\documentclass[10pt,aps,prl,twocolumn,superscriptaddress]{revtex4}

\usepackage{graphics}
\usepackage{epsfig}
\usepackage{dcolumn}
\usepackage{bm}
\usepackage{amsmath}
\usepackage{dsfont} 
 \usepackage{mathtools}
 \usepackage{amsfonts}
\usepackage{latexsym}
\usepackage{amssymb}
\usepackage{color}
  \usepackage{hyperref}
\usepackage[normalem]{ulem}
 \usepackage{bbm}
\usepackage{bbold}

 \newcommand{\tb}[1]{}

\newcommand{\bs}{\boldsymbol}

\newcommand{\g}[1]{{\bf #1}}
\newcommand{\stkout}[1]{\ifmmode\text{\sout{\ensuremath{#1}}}\else\sout{#1}\fi}

\newcommand{\be}{\begin{equation}}
\newcommand{\ee}{\end{equation}}
\newcommand{\bea}{\begin{eqnarray}}
\newcommand{\eea}{\end{eqnarray}}
\newcommand{\ba}{\begin{eqnarray*}}
\newcommand{\ea}{\end{eqnarray*}}

\begin{document}

\title{Berry-phase induced entanglement of hole-spin  qubits in a microwave cavity}

  \author{Marcin M. Wysoki\'nski}
  \email{wysokinski@magtop.ifpan.edu.pl}
  \author{Marcin P\l odzie\'n}
  \author{Mircea Trif}
 \email{mtrif@magtop.ifpan.edu.pl}
  
    \affiliation{International Research Centre MagTop, Institute of
   Physics, Polish Academy of Sciences,\\ Aleja Lotnik\'ow 32/46,
   PL-02668 Warsaw, Poland}
 
\begin{abstract}
    Hole-spins localized in semiconductor structures, such as quantum dots or defects,  serve to the realization of  efficient gate-tunable  solid-state quantum bits. Here we study two electrically driven spin $3/2$ holes coupled to the electromagnetic field of a microwave cavity. We show that the interplay between the non-Abelian Berry phases generated by local time-dependent electrical fields and the shared cavity photons allows for fast manipulation, detection, and long-range entanglement of the hole-spin qubits in the absence of any external magnetic field. Owing to its geometrical structure, such a scheme is more robust against external noises than the conventional hole-spin qubit implementations. These results suggest that hole-spins are favorable qubits for scalable quantum computing by purely electrical means.
\end{abstract} 
 
\date{\today}
\maketitle 
\emph{Introduction}.— Spin-based solid state quantum bits (qubits) are among the most desirable platforms for implementing a quantum processor as they are inherently scalable, they interact weakly with the environments, and can be integrated efficiently with  electronics \cite{LossPRA1998,KaneNature98,PettaScience2005,MortonNature08,Awschalom1174,VeldhorstNature2015,KoppensNature06,Nadj-PergeNature2010,HuNatNano12,MuhonenNature14,LilesNatComm18}. 

Electric fields, instead of the conventional magnetic fields, are preferred for quantum manipulation as they can be applied locally, can be made strong, and can be switched on and off fast. Spins in solids, and specifically in semiconductors, can experience strong spin-orbit interactions (SOIs) that allow for coherent electrical spin control. Most of the implementations and proposals rely on the SOI mechanism facilitated by the presence of a static magnetic field that breaks the time-reversal symmetry. However, generating such a coupling purely electrically, without breaking this symmetry would be advantageous as it would deactivate various dephasing mechanisms that rely on charge fluctuations, such as phonons  and gate voltage noise \cite{GolovachPRL04,SerebrennikovPRL04,PabloPRL06,GerardotNature2008,TrifPRL10}. 

A variety of schemes that utilise the  {\it non-Abelian} geometric phase acquired by the spin qubits  states in the presence of SOI and external electrical fields have been proposed for manipulating geometrically spins in solid state devices without the need for an applied magnetic field \cite{bernevigPRB05,PabloPRB08,GolovachPRA10,BudichPRB12}. Of particular interest are the hole-spin qubits realized in the $S=3/2$ valence band of many semiconductors \cite{bernevigPRB05,BudichPRB12}. They posses strong SOI, and the $p$-type character of the orbital wave-functions leads to a suppression of the hyperfine coupling to the surrounding nuclei \cite{FischerPRB2008}. Experimentally, hole-spins have been under intense scrutiny recently \cite{LossPRL2005,HeissPRB2007,GerardotNature2008,FischerPRB2008,BrunnerScience2009,TrifPRL10, WarburtonNatMat2013,HigginbothamNanoLet2014}, and a lot of progress have been made implementing  conventional one- and two-qubit gates \cite{ WarburtonNatMat2013,HigginbothamNanoLet2014,RoggeNano2014,WatzingerNatCom2018,MourikNature2020,HendrickxNature2020}.  Building on the original works by Avron et al. \cite{AvronPRL88,avron1989}, in Refs.~\cite{bernevigPRB05} and \cite{BudichPRB12} it has been shown explicitly how single {\it geometrical} hole-spin qubit gates \cite{ZanardiPLA1999} can be implemented using only electrical fields. However, to the best of our knowledge, leveraging the geometry of the hole-spin states in order to implement two-qubit gates and create entanglement has yet to be demonstrated.
Such geometrical  entanglement is potentially more robust since it is not affected by gate timing errors and various control voltage inaccuracies.

\begin{figure}[t] 
\centering
\includegraphics[width=0.9\linewidth]{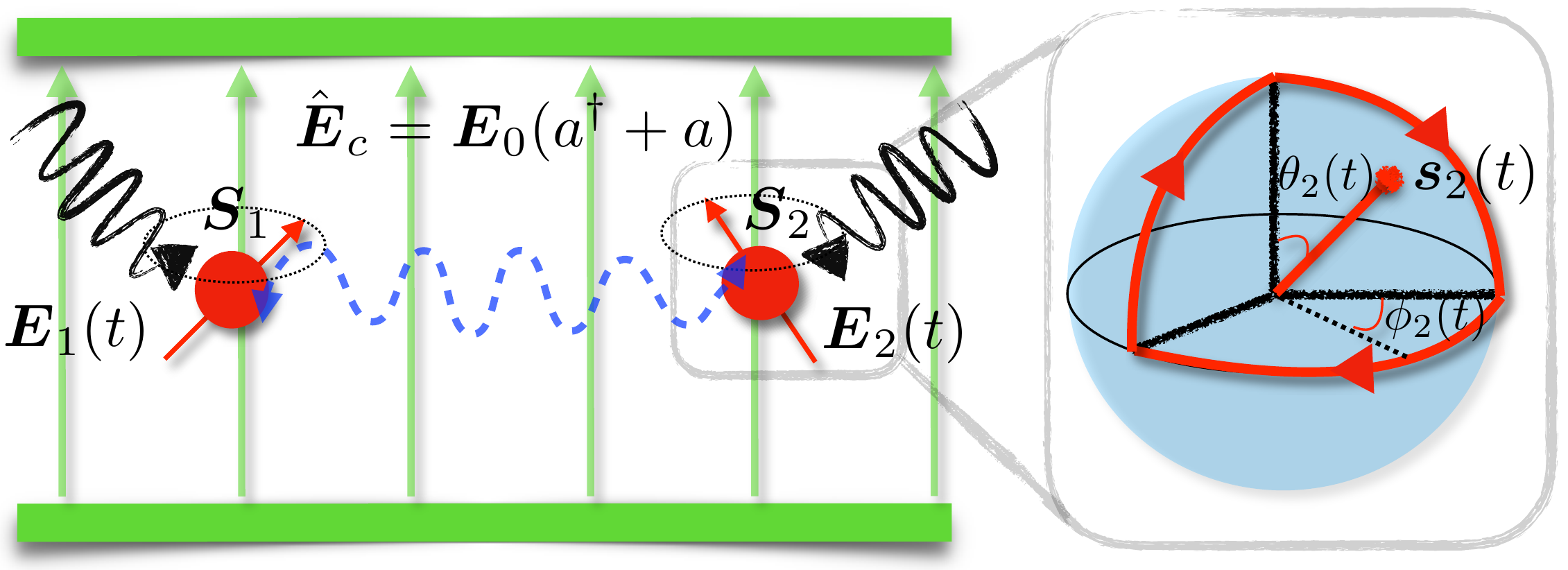} 
\caption{Left: Sketch of the two  hole-spins $S=3/2$ system coupled to a cavity field $\hat{\bs E}_c={\bs E}_0(a^\dagger+a)$. Each of the two spins $j=1,2$ is driven by a classical time-periodic electrical  field ${\bs E}_{j}(t+T_j)={\bs E}_{j}(t)$, with $T_i$ the corresponding period. The cavity induces a time-dependent coupling between the two spins (blue wavy line)  Right: The evolution of one of the effective qubits in the degenerate low-energy sector on the Bloch sphere during the adiabatic driving. Here ${\bs s}_2(t)$ is the instantaneous direction of the effective magnetic field quantified by the angles $\theta_2(t)$ and $\phi_2(t)$, while in red we exemplified one possible cyclic trajectory.} 
\label{fig1}
\end{figure}

In this work we make this step
and propose a novel way to create entanglement between hole-spin qubits utilising their non-Abelian geometric structure, local electric fields, and the photons in a microwave cavity. We show that: $(i)$ the cavity photons become imprinted with the Berry phases generated during the single hole-spin qubit gates, allowing for an efficient non-destructive qubit readout, and $(ii)$ the interplay between photons and the non-Abelian geometry of the states allow for long-range, entangling hole-spin qubits interactions of  geometrical origin. Moreover, such a coupling is only present when both qubits are electrically driven, making it ideal for selectively coupling hole-spins.

\emph{System and Model Hamiltonian.---} We consider the system shown in Fig.~\ref{fig1}, which consists of two {\it electrically} driven spin $3/2$ coupled to the electric field of a microwave cavity. The minimal Hamiltonian describing the system reads \cite{bernevigPRB05}:
\begin{equation}
   H_{tot}(t)= \sum_{j=1,2}d_j[E_{j,\alpha}(t)+E_{0,\alpha}(a^\dagger+a)\,]\Gamma_j^\alpha+\omega_0a^\dagger a\,,
 \label{Hfull}
\end{equation}
where $d_j$ is the spin-electric field coupling strength of spin $j=1,2$,  $E_{j,\alpha}(t)$ and $E_{0,\alpha}$ are the $\alpha=x,y,z$ components  of the $j=1,2$ (time-dependent) external and cavity electric field, respectively, while $a$ ($a^\dagger$) are the photon annihilation (creation) operator, with $\omega_0$ being the bare cavity frequency. Also, the matrices $\Gamma^{n}_j$, with $n\in\{1,5\}$, are the generators of the SO(5) Clifford algebra for spin $j$ \cite{bernevigPRB05,SM}. The above Hamiltonian is precisely that of  Ref.~\cite{bernevigPRB05} proposed to process spin $3/2$ valence band impurities in III-V semiconductors, but accounting for a quantum electrical field stemming from the cavity on top of the time-dependent classical drive. 
There, the coupling to the electrical field originates from the linear Stark effect allowed by the diamond $T_d$ symmetry,
as it is the case of acceptor spins in Si  \cite{SalfiPRL2016}. 
In such cases, $d=ea_B\chi$, with  $e$, $a_B$ and $\chi$ being the electron charge, the Bohr radius, and the dimensionless dipolar parameter, respectively  \cite{BIR63,CoishPRB20}.   More complicated terms, such as the quadrupolar couplings \cite{BudichPRB12} can be accounted for within the same framework by extending  the couplings to all the $\Gamma^n$  matrices. For simplicity, in the following we substitute  $d_jE_{0,\alpha}\equiv g_{j,\alpha}$ and take $d_j=1$.
In the presence of an electric field the time reversal symmetry is preserved and each spin $3/2$ is described by two doubly degenerate instantaneous
states corresponding to the energies $\pm\epsilon_{1,2}(t)$.

\emph{Adiabatic perturbation theory.---} For static external fields, and in the absence of the cavity,  the spectrum  consists of (at least) double degenerate levels, consequence of the Kramers theorem. In the adiabatic limit, quantified by  $\dot{E}_{j,\alpha}/E_{j,\alpha}\ll2\epsilon_j$, with $2\epsilon_j$ being the instantaneous spin splitting of hole $j$, as well as for weak spin-photon coupling  $|{\bs g}_j|\ll|\epsilon_j-\omega_0|$, we can treat both the dynamics and the coupling to photons in time-dependent perturbation theory. In the following,  we extend the approach in Ref.~\cite{SnizhkoPRB19} used to single out the geometrical effects in degenerate systems to the $S=3/2$ spin system. In contrast to Ref.~\cite{SnizhkoPRB19}, however,  we treat the environment (cavity photons) on the same footing with the two spins $3/2$. The full technical details are left for the supplementary material (SM)\cite{SM}, while here we only describe the steps and summarize the results. That entails to first performing a time-dependent unitary transformation, $U(t)=U_1(t)U_2(t)$, that diagonalises each of the isolated spin $3/2$ Hamiltonian, so that $\widetilde{H}_{tot}(t)=\omega_0n_{ph}+\sum_j[H_{j,0}(t)+V_j(t)]$, where $H_{j,0}(t)=\epsilon_j(t)\Gamma_j^5$ is the unperturbed part of the spin $j=1,2$ Hamiltonian \cite{SnizhkoPRB19}, with $\epsilon_j=\sqrt{\sum_\alpha E^2_{j,\alpha}(t)}$, $n_{ph}\equiv a^\dagger a$,  and
\begin{align}
    \!\!\!V_{j}(t)=\dot{E}_{j,\alpha}\mathcal{A}_{j,\alpha}+g_{j,\alpha}(\partial_{\alpha}\epsilon_j\,\Gamma_j^5+i\epsilon_j[\mathcal{A}_{j,\alpha},\Gamma^5_j])X_{ph}\,.
\end{align}
Here $\mathcal{A}_{j,\alpha}= -iU^\dagger_j(t)\partial_{E_{j,\alpha}}U_j(t)$ is the non-Abelian gauge field pertaining to the electric field $E_{j,\alpha}$ with $\partial_\alpha\equiv\partial_{E_{j,\alpha}}$, and $X_{ph}=(a^\dagger+a)$.
Note that $V_j(t)$ leads to both diagonal and off-diagonal transitions between the degenerate eigenstates of the bare spin Hamiltonian $H_{j,0}(t)$. Next we can treat the $\dot{E}_{j,\alpha}$ and $V_j(t)$ in perturbation theory with respect to the spin splittings $\epsilon_j$ and photon frequency $\omega_0$  using a time-dependent Schrieffer-Wolff (SW) transformation $U'(t)=U_1'(t)U_2'(t)$, with  $U_j'(t)=e^{-S_j(t)}\approx1-S_j(t)+S_j^2(t)/2+\dots$. By imposing  $[S_j(t),H_{j,0}+\omega_0a^\dagger a]+V_j(t)=0$, it allows us to keep the leading diagonal terms in the velocities $\dot{E}_{j,\alpha}$ and the second order corrections in $g_{j,\alpha}$. Then,  projecting onto the low four-dimensional energy subspace spanned by the  $\{-\epsilon_{1},-\epsilon_2\}$,  we can find an explicit expression for $S_j(t)$ (see SM for details). That in turn allows us to obtain the low-energy spin-photon Hamiltonian $\delta\mathcal{H}(t)=\sum_j\delta\mathcal{H}_j(t)+\mathcal{H}_{1-2}(t)$, with 
\begin{align}
    \!\!\!\delta\mathcal{H}_j(t)&=\dot{E}_{j,\alpha}g_{j,\beta}(\mathcal{F}^l_{j,\alpha\beta}X_{ph}+g_{j,\gamma}\mathcal{O}^{l}_{j,\alpha\beta\gamma}n_{ph})\nonumber\,,\\
    \mathcal{H}_{1-2}(t)&=\frac{2g_{1,\alpha}g_{2,\beta}}{\omega_0}\dot{E}_{1,\gamma}(t)\dot{E}_{2,\delta}(t)\mathcal{F}^l_{1,\alpha\gamma}\mathcal{F}^l_{2,\beta\delta}\,,
\label{effHam}    
\end{align}
representing the photon-dependent single hole-spin Hamiltonian and the cavity-mediated spin-spin coupling term, respectively. Here, $\mathcal{A}^l_{j,\alpha}\equiv\mathcal{P}_j^l\mathcal{A}_{j,\alpha}\mathcal{P}_j^l$, with $\mathcal{P}_{j}^l$ a projector onto the low-energy degenerate subspace of spin $j$, $\mathcal{F}^l_{j,\alpha\beta}=\partial_\alpha\mathcal{A}^l_{j,\beta}-\partial_\beta\mathcal{A}^l_{j,\alpha}+i[\mathcal{A}^l_{j,\alpha},\mathcal{A}^l_{j,\beta}]$ is the corresponding non-Abelian Berry curvature, and  
 $\mathcal{O}^{l}_{j,\alpha\beta\gamma}$ is an operator that encode also the geometry of the states. In particular, for $\omega_0\ll\epsilon_{1,2}$, this can be written as
\begin{align}
    \mathcal{O}^{l}_{j,\alpha\beta\gamma}&=i[\partial_\alpha \mathcal{A}_{j,\beta},\mathcal{A}_{j,\gamma}]^l-2\partial_\beta\log[\epsilon_j]\mathcal{F}^l_{j,\gamma\alpha}\nonumber\\
    &-2\left(\mathcal{G}^l_{j,\beta\gamma}\mathcal{A}^l_{j,\alpha}-\mathcal{A}^-_{j,\beta}  \mathcal{A}^h_{j,\alpha}\mathcal{A}^+_{j,\gamma}\right)\,,
\end{align}
where $[\dots]^l\equiv\mathcal{P}_j^l[\dots]\mathcal{P}_j^l$, $\mathcal{G}^l_{j,\beta\gamma}$ is the quantum metric in the lowest subspace \cite{SM}, and $\mathcal{A}_{j,\alpha}^h\equiv\mathcal{P}_j^h\mathcal{A}_{j,\alpha}\mathcal{P}_j^h$, with $\mathcal{P}_j^h=1-\mathcal{P}_j^l$ being the Berry curvature in the highest energy subspace of spin, and $\mathcal{A}_{j,\alpha}^{+(-)}\equiv\mathcal{P}_j^{h(l)}\mathcal{A}_{j,\alpha}\mathcal{P}_j^{l(h)}$. The  Hamiltonians in Eq.~\ref{effHam} are the central results of this work, showing that photons in a cavity can be imprinted with the individual hole-spin Berry phases and, moreover, they can mediate interactions between two hole-spins via the geometry of their states  in the absence of any external magnetic fields. Therefore, such effects are present only if the spins are driven, providing means for selectively entangling spin $3/2$ qubits coupled to the same cavity field. Notably, the above Hamiltonians depend only on the geometry of states through their Berry connections, being thus general and applicable, we believe, to any non-Abelian system.  Although the Hamiltonian $\mathcal{H}_{1-2}(t)\propto\dot{E}_{1,\gamma}(t)\dot{E}_{2,\delta}(t)$, the evolution operator endowed by this term is effectively geometrical when the two driving frequencies $\Omega_{1,2}$ (in a continuous operation mode) are incommensurate \cite{MartinPRX17,SM}, as we show explicitly later.

The first term in $\delta\mathcal{H}_{j}(t)$ in Eq.~\ref{effHam} describe  the leading order coupling of the degenerate spin $3/2$ subspace to the photons, in agreement with the findings in Ref.~\cite{SnizhkoPRB19}. This term can be leveraged in order to manipulate the qubit by driving the cavity with a classical (coherent) field. The second contribution instead is novel and accounts for the cavity frequency shift induced by the individual hole-spins geometry of states. Thus, we have extended the dispersive read-out of geometrical Abelian Berry phases \cite{KohlerPRL17,TrifPRL19} to the non-Abelian realm.  While seemingly complicated, the origin of each term in $\mathcal{O}_{j,\alpha\beta\gamma}^{l}$ can be  unravelled by using a Floquet approach for describing the dynamics \cite{SM}. Interestingly, for $\omega_0\sim\dot{E}_{j,\alpha}/\epsilon_j$, the photons and the external driving become resonant, and given that generally $[\mathcal{A}^l_{j,\alpha}, \mathcal{F}^l_{j,\alpha\beta}]\neq0$, it can result in a novel type of Jaynes-Cummings Hamiltonian that is activated by the geometry of the states. Nevertheless, we leave this aspect for future work, and focus here on the regime $\omega_0\gg\dot{E}_{j,\alpha}/\epsilon_{j}$.    

\emph{Dispersive Floquet approach.---} Next we utilise a Floquet description of the hole-spins dynamics that is appropriate when each of the spins $3/2$ is driven periodically, or
$H_{j}(t+T_j)=H_{j}(t)$ ($H_{j}(t)\!\!\equiv\!\!E_{j,\alpha}(t)\Gamma_j^\alpha$),
with $\Omega_j\!\!=\!\!2\pi/T_j$ being the driving frequency of spin $j=1,2$. In the absence of the cavity, the time-dependent wave-functions (or Floquet states) can be written as $|\Psi_j^{s}(t)\rangle=e^{-i\mathcal{E}_j^s t}|\psi_j^s(t)\rangle$, where $|\psi_j^s(t+T_j)\rangle=|\psi_j^s(t)\rangle$ is found as solutions to the Schr\"odinger equation $\mathcal{H}_{j}(t)|\psi_j^s(t)\rangle\equiv[H_{j}(t)-i\partial/\partial t]|\psi_j^s(t)\rangle=\mathcal{E}_j^s|\psi_j^s(t)\rangle$, and $\mathcal{E}_j^s$  are the Floquet eigenvalues for spin $j$ that are defined up to multiple of $\Omega_j$, with $s=1,2,\dots$ labelling the periodic  Floquet states.  In the adiabatic limit $\mathcal{E}_j^s=\epsilon_j^s+ \gamma_j^s/T_j$, with $\epsilon_j^s$ and $\gamma_j^s$ being the instantaneous (or average) energy and the Berry phase of the spin $j$ in the Floquet state $s$. Coupling the spins to the photons results in both shifts in the individual Floquet energies and a coupling between the two spins. The full dynamics of the two spins driven at different frequencies is rather involved (see, for example, Ref.~\cite{MartinPRX17}), and  here instead we focus on the weak coupling regime in the dispersive limit.  That is when $|\Delta_j^{ss'}(q)-\omega_0|\gg |{\bs g}_{1,2}|$, with $\Delta_j^{ss'}(q)=|\mathcal{E}_j^{s}-\mathcal{E}_j^{s'}-q\Omega_j|$ and $q\in\mathcal{Z}$, which allows us to treat the spin-photon interaction in perturbation theory. Using a time-dependent SW transformation, which is described in detail in the SM,  the cavity induced low (quasi-)energy spin Hamiltonian can be cast as $\delta\mathcal{H}=\sum_j\delta\mathcal{H}_{j}+\mathcal{H}^z_{1-2}+\mathcal{H}^\perp_{1-2}$, with  
\begin{align}
\delta\mathcal{H}_{j}&=n_{ph}\sum_{q,s,s'}(-1)^s|V^{ss'}_j(q)|^2\frac{\Delta_j^{ss'}(q)}{[\Delta_j^{ss'}(q)]^2-\omega_0^2}\sigma_j^z\,,\nonumber\\
\mathcal{H}_{1-2}^z&=\frac{2}{\omega_0}\sum_{j,s,p\in low}(-1)^{s+p}V_j^{ss}(0)V_{\bar{j}}^{pp}(0)\sigma^z_1\sigma^z_2\,,\label{Floquet}\\
\mathcal{H}_{1-2}^\perp&=\sum_{j}V_j^{12}(0)V_{\bar{j}}^{21}(0)\frac{2\omega_0}{\omega_0^2-[\Delta_j^{12}(0)]^2}\sigma^+_1\sigma^-_2+{\rm h. c.}\nonumber\,,
\end{align}
where $V_{j}^{ss'}(q)=(1/T_j)\int_0^{T_j}dte^{-iq\Omega_jt}\langle\psi_{j}^s(t)|{\bs g}_j\cdot{\bs \Gamma}_j|\psi_j^{s'}(t)\rangle$ are the Fourier components of the spin-photon matrix elements between states $s$ and $s'$ and spin $j=1,2$. Also, $\sigma_j^\alpha$, with $\alpha=x,y,z$ are Pauli matrices acting in the two lowest (quasi-)energy Floquet states of the hole-spin $j=1,2$. The first term leads to a cavity frequency shift that depends on the  Floquet state of spin $j$, while the second and third terms account for an Ising and $XY$ couplings between the lowest spin Floquet doublets, respectively. As showed in detail in the SM, in the adiabatic limit $\Omega_j\ll|{\bs E}_j|$ we find that $\delta\mathcal{H}_{j}\propto\Omega_j$ and $\mathcal{H}_{1-2}^{z,\perp}\propto\Omega_1\Omega_2$, consistent with the expressions found in the previous section. Note that $\mathcal{H}_{1-2}^{z,\perp}$ depend only on the $q=0$ Fourier components of $V_j^{ss'}(t)$ which, as argued before, result in geometrical effects only on the evolution. All these effects are absent in the static case and, in particular, the entanglement between the  Floquet states is ignited only by driving {\it both} spins.

\emph{Circular driving.---} In order to verify both the adiabatic theory and the above Floquet approach, in this section we consider a specific model, namely that of a circularly driven spin $3/2$. Without loss of generality in the following we shall use parametrization  ${\bs n}_j(t)=\{-\sin\theta_j\sin\Omega_j t,\sin\theta_j\cos\Omega_j t,\cos\theta_j \}$,  where $\Omega_j$ and $\theta_j$ are again the driving frequency and the cone angle for the $j$-th spin. We were able to find a time dependent transformation $\tilde U(t)$ (for details see SM) that makes the bare hole-spin part of $H_{tot}(t)$ fully time-independent and diagonal; $i.e.$ it gives access to the exact solution in the absence of the cavity. Therefore, the entire time-dependence of the spins-photon system in this new frame  is shifted to the spin-photon interactions. Then, in the dispersive regime  we can decouple the spin and photonic degrees of freedom by means of the second order SW transformation in ${\bs g}_{j}$ the resulting low-energy spin-photon Hamiltonian assumes the same form as in Eq.~\ref{Floquet} with $\mathcal{H}_{1-2}^\perp=0$. In general  $\delta\mathcal{H}_j=\delta\omega^g_{0,j}(t)\,\sigma_j^z\,n_{ph}$ and $\mathcal{H}_{1-2}^z(t)=J_{1-2}^z(t)\sigma_z^z\,\sigma_2^z$, with $J_{1-2}^z(t)=-(\Omega_1\Omega_2/2\omega_0)f_1(t)f_2(t)$ and $f_j(t+T_j)=f_j(t)$. For ${\bs g}_j=\{0,0,g_j\}$ and in leading order in $\Omega_j$, we obtain (for the general expressions see SM) 
\begin{align}
        \delta\omega^g_{j,0}&=-\frac{2g_j^2\Omega_j(12\epsilon_j^2-\omega_0^2)\cos\theta_j\sin^2\!\theta_j}{(4\epsilon_j^2-\omega_0^2)^2}\,,\\
        f_j&=\frac{g_j\sin^2\!\theta_j}{\epsilon_j}\,,
        \label{CD}
\end{align}
while $\mathcal{H}_{j}=(1/2)\Omega_j\cos\theta_j\sigma_j^z$ (bare low-energy hole-spin Hamiltonian). Above, $\delta\omega_{j,0}^g$ is the cavity frequency shift pertaining to the geometrical imprints of the lowest energy sector, while we disregarded the (dynamical) contributions $\delta\omega_{j,0}^d$ that can shift the cavity frequency by a value independent of the qubit state \cite{SM}.   

\begin{figure}[t] 
\centering
\includegraphics[trim={0 0.28cm 0 0.2cm},clip,width=0.97\linewidth]{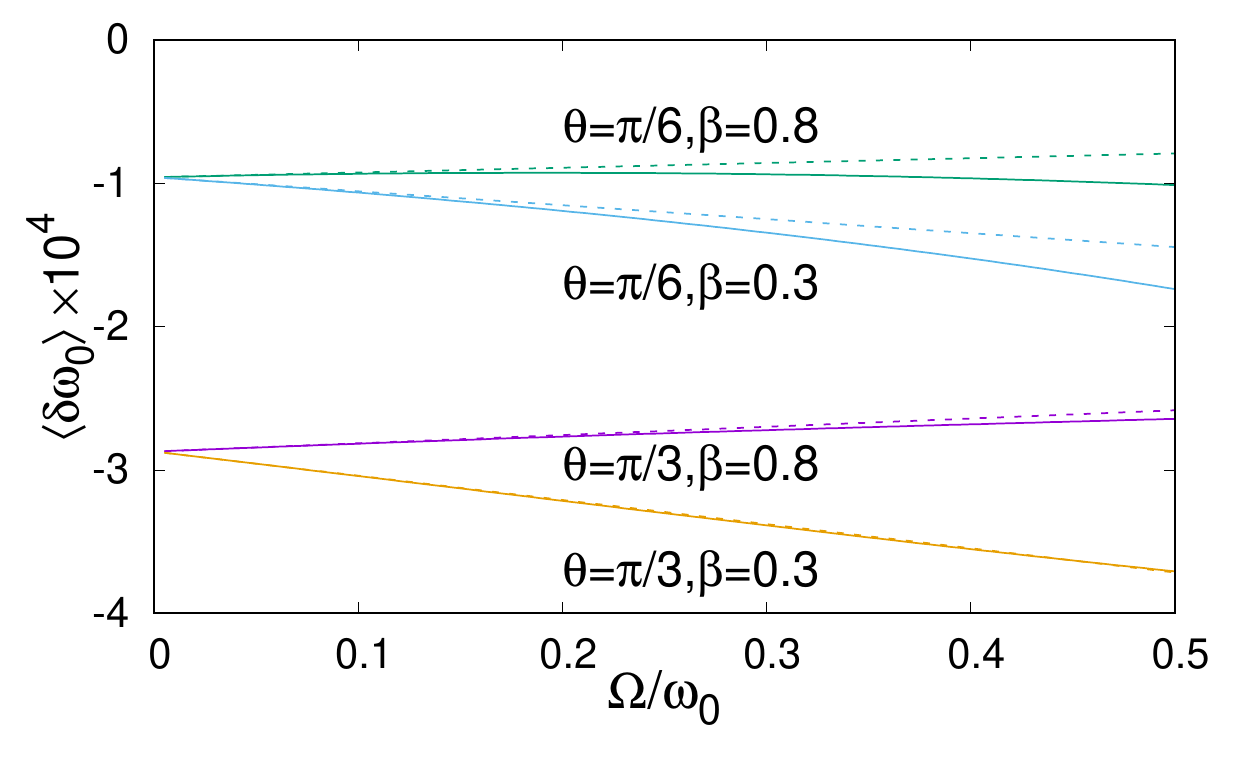}
\caption{Cavity frequency shift $\langle\delta\omega_{0}\rangle$ due to the interaction with a single hole-spin as a function of driving frequency $\Omega$ for several cone angles $\theta$ and initial superposition weights $\beta$. The solid (dashed) lines represent the result without (with) adiabatic approximation. The other parameters  are $\omega_0=0.15$, $\epsilon=1.05$, $g=0.02$, and the spin-photon coupling is set along the $z$ axis.
}
\label{shift} 
\end{figure}

 In the following we demonstrate numerically that in the presence of the driving the cavity frequency shift provides a read-out of the non-Abelian evolution. Given an initial hole-spin state at time $t=0$, $|\psi(0)\rangle=\{\sqrt{1-\beta^2},\beta {\rm e}^{i\phi}\}$, we can evaluate the  geometrical contribution  during the periodic evolution as $\langle\delta\omega^g_{0}\rangle=(1/T)\int_0^{T}\langle\psi(t)|\sigma^z|\psi(t)\rangle\, \delta\omega_{0}^g$, where $|\psi(t)\rangle\equiv \mathcal{U}(t)|\psi(0)\rangle$ with the evolution operator $\mathcal{U}(t)$ describing the bare hole-spin Hamiltonian. In linear order $\Omega$, we find the simple functional dependence $\langle\delta\omega^g_{0}\rangle=(2\beta^2-1)\delta\omega_{0}^g$, which allows to discriminate between different qubit states. As expected, in the absence of the driving $\langle\delta\omega^g_{0}\rangle=0$. In Fig. \ref{shift} we plot the total photonic frequency shift $\langle\delta \omega_{0}\rangle\equiv\langle\delta \omega^d_{0}\rangle+\langle\delta \omega^g_{0}\rangle$ obtained from evolving the full spin $S=3/2$ Hamiltonian and that obtained from the adiabatic, low-energy approximation, respectively as a function of the driving frequency $\Omega$ for various values of $\beta$ \cite{SM}. We see that the adiabatic approximation (linear in $\Omega$) describes well the frequency shift for a wide range of parameters \cite{SM}.
 
 \begin{figure}[t] 
 \includegraphics[width=0.95\linewidth]{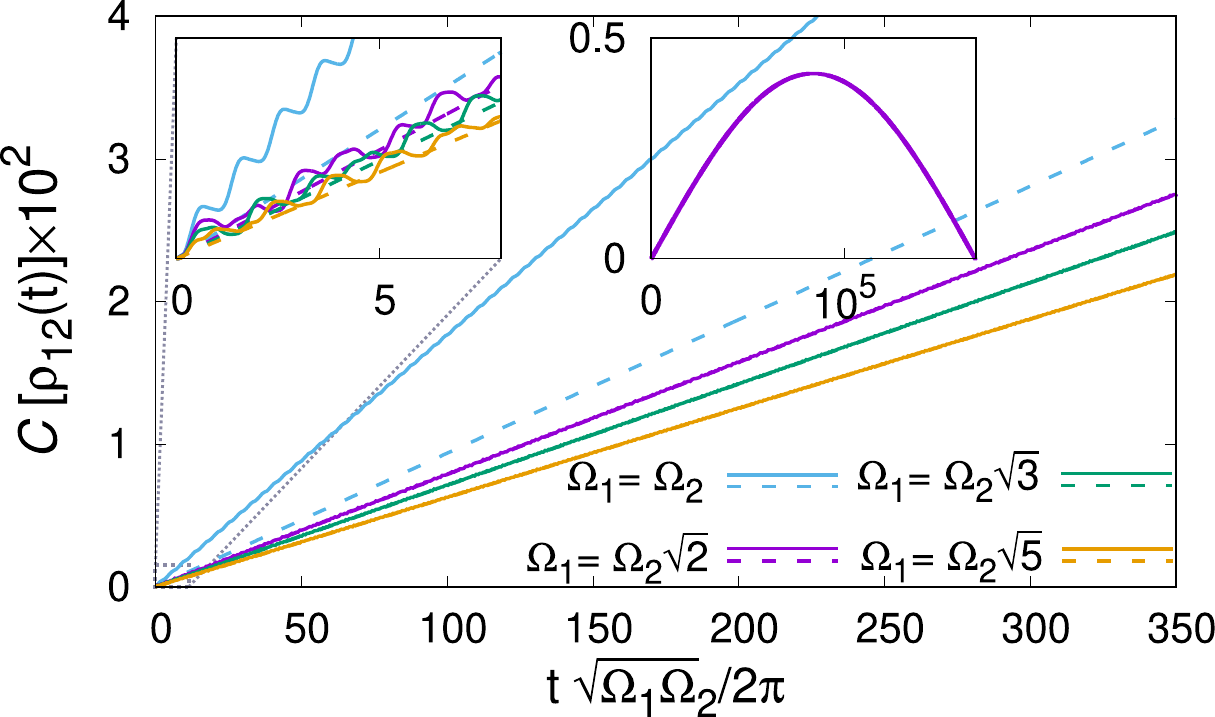} 
 \caption{The concurrence $C[\rho_{12}(t)]$ pertaining to the two-qubit density matrix $\rho_{12}(t)$ as a function of time for various driving $\Omega_1/\Omega_2$ ratios. With dashed lines we mark concurrence generated by  $\mathcal{H}^{z,g}_{1-2}$. The left inset depicts the short time behaviour of the concurrence exhibiting fast oscillations, while the right inset shows a non-monotonic behavior of concurrence for long times $t\simeq\hbar/J_{1-2}^z$ for $\sqrt{2}\Omega_2=\Omega_1$. 
 We have used following parametrization: $\omega_0=0.15$, $g=0.02$, $\epsilon_1=1.05$,   $\epsilon_2=0.95$, $\Omega_1=0.1$, ${\bs g}_1=g\{1/2,1/2,1/\sqrt{2}\}$, ${\bs g}_2=g\{1/\sqrt{2},1/2,1/2\}$, $\theta_1=\pi/3$,  $\theta_2=\pi/4$, $\beta_1=0.4$, and $\beta_2=0.3$.
 }
 \label{con} 
\end{figure}
 
 Finally, we demonstrate the entangling properties of the Hamiltonian $\mathcal{H}^z_{1-2}(t)$. Before that, it is instructive to define an effective static Hamiltonian, $\mathcal{H}_{1-2}^{z,g}=J^{z,g}_{1-2}\sigma^z_1\sigma^z_2$ with  $J^{z,g}_{1-2}=-\Omega_1\Omega_2f_1^0f_2^0/2\omega_0$, where $f_j^0=(\Omega_j/2\pi)\int_0^{2\pi/\Omega_j}d\tau f_j(\tau)$ is the $q=0$ Fourier component of $f_j(t)$. Since each closed spin trajectory is contributing individually here, $\mathcal{H}_{1-2}^{z,g}$ is inherently geometrical (external noises affect its evolution similarly to geometrical gates). For a given two-qubit density matrix, $\rho_{12}(t)$, the entanglement can be quantified by the concurrence  $C[\rho_{12}(t)]={\rm max}[0,\lambda_{12}^1-\lambda_{12}^2-\lambda_{12}^3-\lambda_{12}^4]$ \cite{WootersPRL98} where the $\lambda_{12}^k$ are the eigenvalues of the Hermitian matrix $R_{12}=\sqrt{\sqrt{\rho_{12}}\tilde{\rho}_{12}\sqrt{\rho_{12}}}$ sorted in descending order with $\tilde{\rho}_{12}=(\sigma_1^y\otimes\sigma_2^y)\rho_{12}^*(\sigma_1^y\otimes\sigma_2^y)$. The concurrence is $C=0(1)$ for a separable (maximally entangled) state. Starting from a separable density matrix at $t=0$, in  Fig. \ref{con} we show  $C[\rho_{12}(t)]$ as a function of time when the evolution is generated by the full time-dependent $\mathcal{H}^z_{1-2}(t)$ Hamiltonian and by the effective static Hamiltonian $\mathcal{H}^{z,g}_{1-2}$.  We see excellent (poor) agreement between the two cases when the driving frequencies are incommensurate (commensurate), demonstrating the geometrical origin of the entanglement at incommensurate drives. Note that $C[\rho_{12}(t)]$ increases with time, becoming maximal for $t\sim\hbar/J_{1-2}^z$ (cf. right inset of Fig. \ref{con}). Furthermore, we also analyzed the robustness of the entanglement generation to noises in the driving frequencies, $\Omega_j t \rightarrow \phi_j(t)\equiv \Omega_j t + \delta_j(t)$, with $\delta_j(t)$ being a gaussian correlated noise \cite{MartinPRX17}. We have evaluated  $\kappa = |C_0(t) - \overline{C(t)}|$, where $C_0(t)$ is noiseless concurrence and $\overline{C}(t)$ is the mean concurrence, we found that $\kappa$ is almost two orders of magnitudes smaller in the case of incommensurate drives as compared to the commensurate ones \cite{SM}. That is again consistent with the enhanced protection associated with geometrical qubits \cite{ZanardiPLA1999}.

 In order to give some estimates for the strength of the exchange coupling induced by the dynamics presented in this work,  we utilise the GaAs quantum dot model proposed  in Ref.~\cite{BudichPRB12}. We assume for the hole-spin splittings  $\epsilon_1=\epsilon_2=0.285$ meV (which corresponds to electrical fields in the range of $10^5-10^6$ V/m), $\omega_0\simeq10$ GHz,  driving frequency $\Omega_1=\sqrt{2}\Omega_2=0.043$ THz,  and spin-cavity couplings strengths $g_1=g_2=5.7$ $\mu$eV. For a  cavity field parallel to $z$-axis, the spin-spin interaction is maximized for $\theta_1=\theta_2=\pi/2$, as showed in Eq.~\ref{CD},  and we obtain $J_{1-2}^z\simeq 2.7$ neV, or a two-qubit gate time of $10^{-5}$ s. This time scale is much shorter than the coherence times of a single spin-hole qubit that can be as high as $10$ ms \cite{KobayashiNature2021}.

\emph{Conclusions.---} We have proposed and studied an all-electrical scheme for entangling hole-spins in nanostructures using the non-Abelian character of their states and the electrical field of a microwave cavity. We showed that the Berry phases of electrically driven hole-spins imprint onto the cavity photons allowing for a dispersive readout of the hole-spin qubit. Furthermore, we have shown that the cavity mediates long-range entangling coupling between the non-Abelian Berry curvatures of two hole-spins when both are electrically driven, enabling  selective entanglement between hole-spin qubits. Our work might be relevant for a plethora of other solid-state qubits with non-trivial geometry of states, such as electrons localized in quantum dots or molecular magnets. 

\emph{Acknowledgments.---} This work was supported by the International Centre for Interfacing Magnetism and Superconductivity with Topological Matter project, carried out within the International Research Agendas program of the Foundation for Polish Science co-financed by the European Union under the European Regional Development Fund. We would like to thank Silas Hoffman and Alexander Lau for useful discussions.

 \clearpage
\widetext 
\section{\Large Supplemental Material}

\section{Spin-hole Hamiltonian and $\Gamma$ matrices}
Acceptor impurities in semiconductors with strong spin-orbit coupling due to $p$-type cubic symmetry have
two Kramers degenerate bands of spin $S=3/2$ with spin projections $\pm 1/2$ and $\pm 3/2$. 
An externally applied electric field, $\g E\equiv\{E_x,E_y,E_z\}$, interacts the acceptor bound hole state through the linear Stark effect
\begin{equation}
    H =\frac{e a_B\chi}{\sqrt{3}}(E_x\{S_y,S_z\}+E_y\{S_z,S_x\}+E_z\{S_x,S_y\})\,,
\end{equation}
where $S_{j}$, with $j=x,y,z$, are the spin $3/2$ matrices, $\{A,B\}=AB+BA$ and $a_B$ Bohr radius.
The above Hamiltonian can be elegantly manipulated with the help of the $\Gamma$ matrices \cite{bernevigPRB05} below:
\begin{equation}
\begin{split}
 \Gamma^1&=-\sigma_y\otimes\sigma_x,\\
  \Gamma^2&=-\sigma_y\otimes\sigma_y,\\
   \Gamma^3&=-\sigma_y\otimes\sigma_z,\\
   \Gamma^4&=\sigma_x\otimes\mathbb{1}_2,\\
   \Gamma^5&=\sigma_z\otimes\mathbb{1}_2,\\
\end{split}
 \end{equation}
 satisfying the SO(5) Clifford algebra, i.e. $\Gamma^a\Gamma^b+\Gamma^b\Gamma^a=2\delta_{ab}\mathbb{1}$ and where $\sigma_{x,y,z}$  are the Pauli matrices. The $\Gamma^j$ matrices are related to the spin $3/2$ matrices by
 \begin{equation}
\begin{split}
    &\Gamma^1=\frac{1}{\sqrt{3}}\{S_y,S_z\}\\
    &\Gamma^2=\frac{1}{\sqrt{3}}\{S_z,S_x\}\\
    &\Gamma^3=\frac{1}{\sqrt{3}}\{S_x,S_y\}\\
    &\Gamma^4=\frac{1}{\sqrt{3}}(S_x^2-S_y^2)\\
    &\Gamma^5=S_z^2-\frac{5}{4}\mathbb{1}\\
\end{split}     
 \end{equation}
 The Stark coupling Hamiltonian can then be written simply as:
 \begin{equation}
     H =\g E \cdot \{\Gamma^1,\Gamma^2,\Gamma^3\}\,,
 \end{equation}
 where for the simplicity we have absorbed the constant factors into the electric field vector definition. 
 
\section{Berry connection, Berry curvature, quantum metric tensor}

 For the driven spin $3/2$ Hamiltonian discussed in the Main Text we obtain the following expressions for the Berry connection, Berry curvature, and the metric tensor, respectively, that  act in the $s=low, high$ two-dimensional energy subspace:
 \begin{align}
     \bs{\mathcal A}^s&\equiv-i\mathcal{P}_s\underbrace{U^\dagger \partial_{\bs E}U}_{{\bs A}}\mathcal{P}_s=-\frac{1}{2\epsilon}{\bs n}\times{\bs \sigma}\,,\\
     \mathcal{F}_{\alpha\beta}^s&=i\mathcal{P}_s[A_\alpha, A_\beta]\mathcal{P}_s+i[\mathcal{A}_\alpha^s, \mathcal{A}_\beta^s]=\partial_{E_\alpha}\mathcal{A}_\beta^s-\partial_{E_\beta}\mathcal{A}_\alpha^s+i[\mathcal{A}_\alpha^s,\mathcal{A}_\beta^s]\,,\\
     \bs{\mathcal{B}}^s&\equiv\left(\mathcal{F}_{yz}^s, \mathcal{F}_{zx}^s,\mathcal{F}_{xy}^s\right)=-\frac{1}{2\epsilon^2}(\bs{n}\cdot{\bs \sigma})\,{\bs n}\,,\\
     \mathcal{G}^s_{\alpha\beta}&\equiv\frac{1}{2}\left(\mathcal{P}_s\{A_\alpha,A_\beta\}\mathcal{P}_s-\{\mathcal{A}_\alpha^s,\mathcal{A}_\beta^s\}\right)\Rightarrow\left(
     \begin{array}{cccc}
     \mathcal{G}^s_{xx} & \mathcal{G}^s_{xy} & \mathcal{G}^s_{xz}\\
     \mathcal{G}^s_{yx} & \mathcal{G}^s_{yy} & \mathcal{G}^s_{yz}\\
     \mathcal{G}^s_{zx} & \mathcal{G}^s_{zy} & \mathcal{G}^s_{zz}\\
     \end{array}
\right)=\frac{1}{4\epsilon^2}\left(
     \begin{array}{cccc}
     1-n_x^2 & -n_xn_y & -n_xn_z\\
     -n_xn_y & 1-n_y^2 & -n_yn_z\\
     -n_xn_z & -n_yn_z & 1-n_z^2\\
     \end{array}
\right)\,,     
 \end{align}
where $|{\bs n}|=1$.  
 
\section{Adiabatic perturbation theory}

Here we provide the details on the derivation of the effective low-energy Hamiltonian for the two spins coupled to the cavity. After the first unitary (time-dependent) transformation, the total Hamiltonian can be written as \cite{SnizhkoPRB19}:
\begin{align}
    \widetilde{H}_{tot}=\sum_{j=1,2}\left(\epsilon_j\Gamma^5_j+\dot{E}_{j,\alpha}A_{j,\alpha}+g_{j,\alpha}(\partial_{\alpha}\epsilon_j\,\Gamma_j^5+i\epsilon_j[A_{j,\alpha},\Gamma^5_j])(a^\dagger+a)\right)+\omega_0\,a^\dagger a\,,
\end{align} 
 such that the {\it instantaneous} spin Hamiltonian  is now diagonal, with $\epsilon_j$ being the eigen-energy, possibly still time-dependent. 
 
Next we account for the terms  $\sim\dot{E}_{j,\alpha}$  by diagonalising the spin Hamiltonian in second order in these velocities, at the expense of introducing new coupling terms between the spins and the photons. To achieve that, we perform a unitary transformation on each spin $U^{(2)}_j(t)=e^{-S_j}=1-S_j+(S_j)^2/2+\dots$, with $j=1,2$ and $S_j=-S_j^\dagger$ chosen such that:
\begin{align}
    \dot{E}_{j,\alpha}(1-\mathcal{P}_j)A_{j,\alpha}+\epsilon_j[S_j, \Gamma^5_j]=0\,,
\end{align}
where $\mathcal{P}_jO=\mathcal{P}^h_jO\mathcal{P}_j^h+\mathcal{P}^l_jO\mathcal{P}^l_j\equiv\mathcal{O}^d$ and $(1-\mathcal{P}_j)O=\mathcal{P}^h_jO\mathcal{P}^l_j+\mathcal{P}^l_jO\mathcal{P}^h_j\equiv\mathcal{O}^++\mathcal{O}^-$.
That in turn leads to the following Hamiltonian:
\begin{align}
    \bar{H}_{tot}&=\sum_{j=1,2}\left(\epsilon_j\Gamma^5_j+\dot{E}_{j,\alpha}\mathcal{A}^d_{j,\alpha}+g_{j,\alpha}\left(\partial_{\alpha}\epsilon_j\,(\Gamma_j^5+[S_j,\Gamma_j^5])+i\epsilon_j([A_{j,\alpha},\Gamma^5_j]+[S_j,[A_{j,\alpha},\Gamma^5_j]])\right)(a^\dagger+a)\right)+\omega_0a^\dagger a\,,
\end{align}
where the spins Hamiltonian are diagonalized in leading order in velocities. We then simply obtain:
\begin{equation}
S_j(t)=\frac{\dot{E}_{j,\alpha}}{2\epsilon_j}(\mathcal{A}_{j,\alpha}^+-\mathcal{A}_{j,\alpha}^-)\,, 
\end{equation}
where $\mathcal{A}_{j,\alpha}^{\pm}=\mathcal{P}_j^{h,l}A_{j,\alpha}\mathcal{P}_j^{l,h}$ are the off diagonal raising/lowering type operators stemming from the full gauge field $A_{j,\alpha}$.  With this, the Hamiltonian becomes
\begin{align}
    \bar{H}_{tot}&=\sum_{j=1,2}\epsilon_j\Gamma^5_j+\dot{E}_{j,\alpha}\mathcal{A}^d_{j,\alpha}+\underbrace{g_{j,\beta}\left(\partial_{\beta}\epsilon_j\Gamma_j^5+\dot{E}_{j,\alpha}\mathcal{F}_{j,\alpha\beta}\right)(a^\dagger+a)}_{V_{1,j}(t)}+\omega_0a^\dagger a\nonumber\\
    &-\underbrace{g_{j,\alpha}\left(\frac{\dot{E}_{j,\beta}\partial_{\alpha}\epsilon_j}{\epsilon_j}(\mathcal{A}^+_{j,\beta}+\mathcal{A}^-_{j,\beta})+2i\epsilon_j(\mathcal{A}^+_{j,\alpha}-\mathcal{A}^-_{j,\alpha})\right)(a^\dagger+a)}_{V_{2,j}(t)}\,.
\end{align}
The above Hamiltonian contains explicitly the effective coupling between the photons and the velocity of  spins, while the spin Hamiltonians themselves are now diagonal.    

A second SW transformation, $U_j^{(3)}=e^{-S'_j}$, with $S'_j=-(S'_j)^\dagger$  diagonalizes both the photons and the spins in leading order in the velocities and the spin-photon coupling strength, respectively. In this order, we obtain
\begin{align}
    S'_j&=\frac{g_{j,\alpha}}{\omega_0}\left(\partial_{\alpha}\epsilon_j\Gamma_j^5+\dot{E}_{j,\beta}\mathcal{F}_{j,\beta\alpha}\right)(a^\dagger-a)-2ig_{j,\alpha}\epsilon_j\left[\left(\frac{1}{2\epsilon_j-\omega_0}a+\frac{1}{2\epsilon_j+\omega_0}a^\dagger\right)\mathcal{A}^+_{j,\alpha}+\left(\frac{1}{2\epsilon_j+\omega_0}a+\frac{1}{2\epsilon_j-\omega_0}a^\dagger\right)\mathcal{A}^-_{j,\alpha}\right]\nonumber\\
    &-\frac{g_{j,\alpha}\dot{E}_{j,\beta}\partial_{\alpha}\epsilon_j}{\epsilon_j}\left[\left(\frac{1}{2\epsilon_j-\omega_0}a+\frac{1}{2\epsilon_j+\omega_0}a^\dagger\right)\mathcal{A}^+_{j,\beta}-\left(\frac{1}{2\epsilon_j+\omega_0}a+\frac{1}{2\epsilon_j-\omega_0}a^\dagger\right)\mathcal{A}^-_{j,\beta}\right]\,,
\end{align}
which then leads for the effective Hamiltonian (keeping only the diagonal terms in the leading order in velocities and second order in $g_{j,\alpha}$):
\begin{align}
    \bar{H}_{tot}&=\sum_{j=1,2}\epsilon_j\Gamma^5_j+\dot{E}_{j,\alpha}(\mathcal{A}^d_{j,\alpha}+\frac{1}{2}[S_{p}',[S_k',\mathcal{A}^d_{j,\alpha}]])+\frac{1}{2}[S_p',V_{1,j}]+\frac{1}{2}[S_p',V_{2,j}]-\frac{i}{2}(\dot{S}_j'S_p'-S_j'\dot{S}_p')+\omega_0a^\dagger a\,,
\end{align}
where we neglect all the terms are off-diagonal and lead to higher orders than those accounted for in the following. From above, we can  obtain the single spin coupling Hamiltonians pertaining to the low-energy sector as follows:
\begin{align}
    \delta\mathcal{H}_{j}&=\frac{4\dot{E}_{j,\alpha}g_{j,\beta}g_{j,\gamma}\epsilon_j}{(2\epsilon_j)^2-\omega_0^2}\left[\epsilon_j\frac{(2\epsilon_j)^2+\omega_0^2}{(2\epsilon_j)^2-\omega_0^2}\left[i[\partial_\alpha \mathcal{A}_{j,\beta},\mathcal{A}_{j,\gamma}]-2(\mathcal{G}^l_{j,\beta\gamma}\mathcal{A}^l_{j,\alpha}-\mathcal{A}^-_{j,\beta}  \mathcal{A}^h_{j,\alpha}\mathcal{A}^+_{j,\gamma})\right]-2\partial_\beta\epsilon_j\mathcal{F}^l_{j,\gamma\alpha}\right]a^\dagger a\,.
    \label{AdFull}
\end{align}
which, in the limit of small cavity frequency $\omega_0\ll\epsilon_j$ reduces to the expression showed in the main text. 
Above we only kept the terms that depend on the photonic field (Stark shift), and disregarded the Lamb shift. Finally, the coupling between the two spins reads (for $\omega_0\gg \dot{E}_{j,\gamma}$):
\begin{align}
     \mathcal{H}_{1-2}&\approx\frac{2g_{1,\alpha}g_{2,\beta}}{\omega_0}\dot{E}_{1,\gamma}\dot{E}_{2,\delta}\mathcal{F}_{1,\alpha\gamma}\mathcal{F}_{2,\beta\delta}\,.
\end{align}

The effective Hamiltonian acting in the $4\times4$ dimensional subspace leads to an evolution that can result in entanglement. To account for that in the adiabatic description, we evaluate the evolution operator in the interaction picture with respect to the single-qubit Hamiltonians stemming from the  non-Abelian dynamics. This reads:
\begin{equation}
    U^I_{1-2}(t,t')=\mathcal{T}e^{-i\int_{t'}^td\tau\mathcal{H}^I_{1-2}(\tau)}\approx1-i\int_{t'}^td\tau\mathcal{H}^I_{1-2}(\tau)+\dots\,,
\end{equation}
where
\begin{equation}
    \mathcal{H}^I_{1-2}(t)=U_1^\dagger(t)U_2^\dagger(t)\mathcal{H}^I_{1-2}(\tau)U_2(t)U_1(t)=\frac{2g_{1,\alpha}g_{2,\beta}}{\omega_0}\dot{E}_{1,\gamma}\dot{E}_{2,\delta}\mathcal{F}^I_{1,\alpha\gamma}(t)\mathcal{F}^I_{2,\beta\delta}(t)\,,
\end{equation}
and $U_i(t)=\mathcal{T}\exp{[-i\int_0^t}d\tau H_i(\tau)]$. Let us next consider a continuous periodic driving of the two qubits along some given trajectory in the parameter space. Then, we can extract a simple form of the evolution operator for each spin as:
\begin{align}
    U_i(t)=\sum_\sigma e^{i\sigma\epsilon_it}|\psi_{i,\sigma}(t)\rangle\langle\psi_{i\sigma}(0)|\,,
\end{align}
where $|\psi_{i,\sigma}(t+T_i)\rangle=|\psi_{i,\sigma}(t)\rangle$ are the corresponding Floquet states for spin $i=1,2$, and $\epsilon_i\propto1/T_i$ is the corresponding Floquet energy associated with the given trajectory. Then, we obtain:
\begin{align}
\mathcal{F}^I_{i,\alpha\gamma}(t)&=\sum_{\sigma\sigma'} e^{i(\sigma-\sigma')\epsilon_i t}\langle\psi_{i\sigma}(t)|\mathcal{F}_{i,\alpha\gamma}(t)|\psi_{i,\sigma'}(t)\rangle|\psi_{i\sigma}(0)\rangle\langle\psi_{i\sigma'}(0)|\equiv \sum_{\sigma\sigma'} e^{i(\sigma-\sigma')\epsilon_i t}\mathcal{F}^{\sigma\sigma'}_{i,\alpha\gamma}(t)|\psi_{i\sigma}(0)\rangle\langle\psi_{i\sigma'}(0)|\nonumber\\
&= \mathcal{F}^{z}_{i,\alpha\gamma}(t)\sigma_z^i+ e^{2i\epsilon_i t}\mathcal{F}^{\uparrow\downarrow}_{i,\alpha\gamma}(t)\sigma_+^i+e^{-2i\epsilon_i t}\mathcal{F}^{\downarrow\uparrow}_{i,\alpha\gamma}(t)\sigma_-^i
\end{align}
where $\mathcal{F}^{z}_{i,\alpha\gamma}(t)=(1/2)(\mathcal{F}^{\uparrow\uparrow}_{i,\alpha\gamma}(t)-\mathcal{F}^{\downarrow\downarrow}_{i,\alpha\gamma}(t))$ and $\mathcal{F}^{\sigma\sigma'}_{i,\alpha\gamma}(t+T_i)=\mathcal{F}^{\sigma\sigma'}_{i,\alpha\gamma}(t)$. Note that the operators $\sigma_{z,+,-}^i$
 act in the Floquet basis of the individual spins. The resulting slow-oscillating Hamiltonian becomes:
 \begin{align}
 \mathcal{H}^I_{1-2}(t)=\frac{2g_{1,\alpha}g_{2,\beta}}{\omega_0}\dot{E}_{1,\gamma}\dot{E}_{2,\delta}\left\{
 \begin{array}{cc}
 \mathcal{F}_{1,\alpha\gamma}^z(t)\mathcal{F}_{2,\beta\delta}^z(t)\sigma_1^z\sigma_2^z\,, & {\rm for} \,\epsilon_1\neq\epsilon_2\\\\
 \mathcal{F}_{1,\alpha\gamma}^z(t)\mathcal{F}_{2,\beta\delta}^z(t)\sigma_1^z\sigma_2^z+\mathcal{F}_{1,\alpha\gamma}^{\uparrow\downarrow}(t)\mathcal{F}^{\downarrow\uparrow}_{2,\delta}(t)\sigma_+^1\sigma_-^2+{\rm h. c.}\,, & {\rm for}\, \epsilon_1=\epsilon_2
 \end{array}
 \right.\,.
 \end{align}
 Note that the two contributions commute, and thus we can treat them separately in the time-ordering. 
Let us evaluate the evolution operator for  $\epsilon_1\neq\epsilon_2$, the other case being addressable similarly.  Then we obtain:
\begin{align}
    U^I_{1-2}(t,t')=\mathcal{T}e^{-i\int_{t'}^td\tau\mathcal{H}^I_{1-2}(\tau)}=e^{-2i(g_{1,\alpha}g_{2,\beta}/\omega_0)\bar{\mathcal{F}}_{1,\alpha\gamma}^z\bar{\mathcal{F}}_{2,\beta\delta}^z\sigma_1^z\sigma_2^zt}\,,
\end{align}
where 
\begin{equation}
\bar{\mathcal{F}}_{i,\alpha\gamma}^z=\frac{1}{T_i}\int_0^{T_i}dt\dot{E}_{i,\gamma}(t)\mathcal{F}_{i,\alpha\gamma}^z(t)\equiv\frac{1}{T_i}\oint dE_{i,\gamma}\mathcal{F}_{i,\alpha\gamma}^z({\bs E_{i}})\,,
\end{equation}
is the average of the diagonal Berry curvature over the corresponding period. Note that in this case the evolution operator only affects the $\{|\uparrow\uparrow\rangle,|\downarrow\downarrow\rangle\}$ subspaces, while the $\{|\uparrow\downarrow\rangle,|\downarrow\uparrow\rangle\}$ remains inert.

\section{Floquet theory for qubits driven in a cavity}

Let us consider again the time-dependent Hamiltonian describing the two spins in the cavity written in the original form (for $d_j=1$):
\begin{align}
H_{\rm tot}(t)&=\omega_0a^\dagger a+\sum_{j=1,2}\left[{\bs E}_j(t)+{\bs g}_j(a^\dagger+a)\right]\cdot{\bs \Gamma}_j\,,
\label{ham_tot}
\end{align} 
where ${\bs g}_j$ is the (vector) coupling strength of the spin  ${\bs \Gamma}_{j}=(\Gamma_j^1, \Gamma_j^2, \Gamma_j^3)$ to the cavity. As opposed to the previous case, here each of the spin $3/2$  is driven {\it periodically} by classical drives ${\bs E}_{j}(t+T_j)={\bs E}_{j}(t)$, with $T_j$ the corresponding driving period.  For the spin $j$ time-periodic Hamiltonian, $H_j(t+T_j)=H_j(t)$ ($H_j(t)\equiv{\bs E}_j(t)\cdot{\bs \Gamma}_j$), the Floquet states can be found as solutions to the Schrodinger equation
\begin{align}
\mathcal{H}_j(t)|\psi^s_j(t)\rangle\equiv[H_j(t)-i\partial/\partial t]|\psi^s_j(t)\rangle=\mathcal{E}^s_j|\psi^s_j(t)\rangle\,,
\end{align}
where $\mathcal{E}^s_j$ are the Floquet eigenvalues that are defined up to multiples of $\Omega_j$, with $s=1,2,\dots$ labelling the periodic  Floquet states, $|\psi^s_j(t+T_j)\rangle=|\psi^s_j(t)\rangle$.  It is instructive to express the spin-photon coupling in the (complete) Floquet basis of the bare driven spins. In the absence of the coupling to the cavity, we label the Floquet eigenstates of the spin $j=1,2$ by $|\psi^s_j(t)\rangle$. Taking into account the photonic state, in the absence of the coupling between the qubits and the photons, a general Floquet state reads:
\begin{equation}
|\Psi_{ss'n}(t)\rangle=|\psi^s_1(t)\rangle\otimes|\psi^{s'}_2(t)\rangle\otimes|n\rangle\,,
\end{equation}
which will be used as basis states and which satisfy:
\begin{align}
\mathcal{H}_0(t)|\Psi_{ss'n}(t)\rangle=(\mathcal{E}^s_1+\mathcal{E}^{s'}_2+n\omega_0)|\Psi_{ss'n}(t)\rangle\, ,
\end{align}
 where $\mathcal{H}_0(t)= H_1+H_2+\omega_0a^\dagger a -i\partial/\partial t$. 
The above Floquet spectrum, for each spin, can be solved by switching to the Fourier space and mapping the time-dependent problem to a static, eigenvalue problem, or: 
\begin{equation}
|\psi_j^{s}(t)\rangle=\sum_qe^{-iq\Omega_j t}|\psi^{s}_{j}(q)\rangle\,,
\end{equation} 
which then can be substituted into the Floquet Hamiltonian to give the following set of linear equations:
\begin{equation}
\sum_{q}[H_{j}(q-q')+n\Omega_j\delta_{qq'}]|\psi^{s}_{j}(q')\rangle=E^s_j|\psi^{s}_{j}(q)\rangle\,,
\end{equation}
where 
$H_{j}(q-q')=(1/T_j)\int_0^{T_j}dte^{-i(q-q')\Omega_j t}H_j(t)$. Note that now the dimension of the extended Hilbert space is infinite, associated with an infinite number of emitted or absorbed photons. While the number of Floquet energies is infinite, they are defined only up to multiples of $\Omega_j$.  Within this formalism, one can now add the perturbations $V_j\equiv{\bs g}_j\cdot{\bs \Gamma}_j(a^\dagger+a)$ to the Hamiltonian and treat them in the framework of time-dependent perturbation theory. We can write:
\begin{align}
    \mathcal{E}_j^s&=\epsilon_j^s+\bar{\phi}_j^s/T_j\,,\\
    \epsilon_j^s&=(1/T_j)\int_0^{T_j}dt\langle\psi_j^s(t)|H_j(t)|\psi_j^s(t)\rangle+\mathcal{O}(1/T_j^2)\,,\\
    \bar{\phi}_j^s&=i\int_0^{T_j}dt\langle\psi_j^s(t)|d/dt|\psi_j^s(t)\rangle=\gamma_j^s/T_j+\mathcal{O}(1/T_j^2)\,,
\end{align}
being the corresponding  average instantaneous energy and  the Aharonov-Anandan phase, respectively, associated with the Floquet level $s$ in spin $j$. In the adiabatic limit discussed here, the latter term becomes the Berry  phase $\gamma_j^s$, and the average energies $\epsilon_j^s$ will become the instantaneous energies.

A general combined Floquet state satisfies:
\begin{align}
\big[\mathcal{H}_0(t)+\sum_jV_j\big]|\Psi_{r}(t)\rangle\equiv\mathcal{H}(t)|\Psi_{r}(t)\rangle=0\,,
\end{align}
where $|\Psi_{r}(t)\rangle$ are the full Floquet  eigenstates with $r$ labelling index of the mixed spin-photonic state.  This eigenvalue equation resembles the static situation and we proceed to solve it perturbation theory in $V_j$, assuming the weak coupling limit to hold, namely $|{\bs g}_{j}|\ll|{\bs E}_j(t)|,\omega_0$. We relate the full Floquet states to the bare ones by a unitary transformation $|\Psi_{r}(t)\rangle=e^{-i(\mathcal{E}_1^s+\mathcal{E}_2^p)t}U(t)|\Psi_{ss'n}(t)\rangle$, with $U(t)=e^{-S(t)}\approx1-S(t)+S^2(t)/2+\dots$ and $S^\dagger(t)=-S(t)$. We then choose $S(t)$ such that it excludes from $V_j(t)$ the terms that are off-diagonal, i.e. couple different photonic states and, but not necessary, couple different Floquet states. Note that $S(t)=S_1(t)+S_2(t)$ and $S_{1,2}(t+T_{1,2})=S_{1,2}(t)$, and we need only to find each of these transformations individually. Keeping the leading order terms in $V_j$, that pertains to the following equation:
\begin{align}
[S_j(t),\mathcal{H}_{j}(t)]+V_j=0\Leftrightarrow [S_j(t),H_{j}(t)]+V_j-i\dot{S}_j=0\,,
\label{SWF}
\end{align}
which leads to:
\begin{align}
\mathcal{H}(t)\approx\mathcal{H}_0(t)+\frac{1}{2}[S(t),V]\,.
\end{align}
Writing $S_{j}(t)=A_j^+(t)a+A_j^-(t)a^\dagger$, from Eq.~\ref{SWF} above we obtain:
\begin{align}
    \langle \psi_j^s(t)|A_j^{\pm}(t)|\psi_j^{s'}(t)\rangle&=\sum_{q}e^{iq\Omega_jt}\frac{V_{j}^{ss'}(q)}{E_j^s-E_j^{s'}-q\Omega_j\mp\omega_0}\,,
\end{align}
where:
\begin{align}
V^{ss'}_j(q)&=\frac{1}{T_j}\int_0^{T_j}dte^{-iq\Omega_j t}\langle\psi_j^{s}(t) |V_j|\psi_j^{s'}(t)\rangle=\sum_{k}\langle\psi_j^{s}(q) |V_j|\psi_j^{s'}(k+q)\rangle\,,
\end{align} 
and $V^{s's}_j(-p)=[V^{ss'}_j(p)]^*$. We can finally put everything together to obtain:
\begin{align}
S_j(t)&=\sum_{q,s,s'}e^{iq\Omega_j t}V^{ss'}_j(q)\left(\frac{1}{\mathcal{E}_j^s-\mathcal{E}_j^{s'}-q\,\Omega_j+\omega_0}a+\frac{1}{\mathcal{E}_j^s-\mathcal{E}_j^{s'}-q\,\Omega_j-\omega_0}a^\dagger\right)\Sigma_{j}^{ss'}(t)\,,
\end{align}
with $\sigma_{j}^{ss'}(t)=|\psi_j^s(t)\rangle\langle\psi_j^{s'}(t)|$. Writing $V_j$ in the Floquet basis too, we arrive at the dispersive  Hamiltonian:
\begin{align}
\mathcal{H}(t)&\approx\sum_{j=1,2}\left[\mathcal{E}_j+b^{z}_j(t)a^\dagger a\right]\sigma^{z}_j(t)+J^{z}_{1-2}(t)\sigma_1^{z}(t)\sigma_2^{z}(t)+(J^{\perp}_{1-2}(t)\sigma_1^{-}(t)\sigma_2^{+}(t)+{\rm h.c.})\,,\\ 
b^{z}_j(t)&=
    \frac{1}{2}\sum_{q,q',s}(-1)^se^{i(q+q')\Omega_j t}V^{ss'}_j(q)V^{s's}_j(q')\left(\frac{\mathcal{E}_j^s-\mathcal{E}_j^{s'}-q\,\Omega_j}{(\mathcal{E}_j^s-\mathcal{E}_j^{s'}-q\,\Omega_j)^2-\omega_0^2}-\frac{\mathcal{E}_j^{s'}-\mathcal{E}_j^{s}-q'\,\Omega_j}{(\mathcal{E}_j^{s'}-\mathcal{E}_j^{s}-q'\,\Omega_j)^2-\omega_0^2}\right)\,,\\
J^{z}_{1-2}(t)&=\sum_{j,q,q'}(-1)^{s+p}e^{i(q\Omega_j+q'\Omega_{\bar{j}})t}V_j^{ss}(q)V_{\bar{j}}^{pp}(q')\frac{2\omega_0}{\omega_0^2-q^2\,\Omega_j^2}\nonumber\\
J^{\perp}_{1-2}(t)&=\sum_{j,q,q'}e^{i(q\Omega_j+q'\Omega_{\bar{j}})t}V_j^{12}(q)V_{\bar{j}}^{21}(q')\frac{2\omega_0}{\omega_0^2-(\mathcal{E}_j^+-\mathcal{E}_j^--q\,\Omega_j)^2}\,,
\end{align}
where $\sigma_j^\alpha$, with $\alpha=x,y,z$ are here Pauli matrices acting in the low-energy Floquet basis states, and $s,p=\pm$ quantify the lowest (quasi-)energy doublets $E_j^{\pm}$ of the two spins. We can simplify further these expression by only considering the time averages of the above couplings, assuming incommensurate driving frequencies. That simply means $q=-q'$ ($q=q'=0$) in the expression for $b_j^z(t)$ ($J^F_{z,\perp}(t)$). We then  finally obtain the expressions showed in the main text:
\begin{align}
b^{z}_j&=
    \sum_{q,s,s'}(-1)^s|V^{ss'}_j(q)|^2\frac{\mathcal{E}_j^s-\mathcal{E}_j^{s'}-q\,\Omega_j}{(\mathcal{E}_j^s-\mathcal{E}_j^{s'}-q\,\Omega_j)^2-\omega_0^2}\,,\\
J^{z}_{1-2}&=\frac{2}{\omega_0}\sum_{j}(-1)^{s+p}V_j^{ss}(0)V_{\bar{j}}^{pp}(0)\,,\nonumber\\
J^{\perp}_{1-2}&=\sum_{j}V_j^{12}(0)V_{\bar{j}}^{21}(0)\frac{2\omega_0}{\omega_0^2-(\mathcal{E}_j^+-\mathcal{E}_j^-)^2}\,.
\label{exchange}
\end{align}

To connect to the adiabatic approximation, next we  perform a series expansion in $\Omega_j$ in the previous Floquet  expressions. Moreover, we collect only single-spin terms that depend on the photon number and which will lead to changes in the photons frequency,  as well as the resulting cavity mediated spin-spin coupling Hamiltonian. For the former, we can write 
\begin{align}
b^{z}_{j}(t)& 
    \approx-\sum_{q,q',s'\in high}e^{i(q+q')\Omega_j t}(-1)^sV^{ss'}_j(q)V^{s's}_j(q')\left(\frac{4\epsilon_j}{(2\epsilon_j)^2-\omega_0^2}+\Omega_j\frac{(2\epsilon_j)^2+\omega_0^2}{\pi[(2\epsilon_j)^2-\omega_0^2]^2}[\gamma_j^s-\gamma_j^{s'}-\pi (q-q')]\right)\,,
\end{align}
where we used that $\epsilon_{j}^{s}=\epsilon_{j}^{p}=\pm\epsilon_{j}$ with $s,p=low/high$ in leading order on the driving frequency $\Omega_j$. 
To make progress, we write the Floquet states as :
\begin{align}
    |\psi_j^s(t)\rangle=|\phi_{j}^s(t)\rangle+\frac{\Omega_j}{\epsilon_j^s-\epsilon_j^p}\sum_{p}A_j^{ps}(t)|\phi_{j}^p(t)\rangle+\mathcal{O}(\Omega_j^2)\,,
\end{align}
where $|\phi_{j}^s(t)\rangle=|\phi_{j}^s(t+T_j)\rangle$ and $A_j^{ps}(t)=A_j^{ps}(t+T_j)$ are the instantaneous eigenstates and the matrix elements pertaining to the dynamical corrections to these states, respectively. The precise form of $A_{j}^{ps}(t)$ can be found using perturbation theory in $\Omega_j$ from the explicit driving trajectory. Note that the instantaneous wave-functions cannot discriminate between the $s$ and $p$ states associated  to a given  (originally) Kramers doublet, thus all matrix elements that couple such states need to be at least  proportional to $\Omega_j$, i.e. beyond the instantaneous description. Specifically, we can write:
\begin{align}
    V_j^{ss'}(t)&\approx \langle\phi_j^s(t)|V_j|\phi_j^{s'}(t)\rangle+\frac{\Omega_j}{2\epsilon_j}\sum_p(A_j^{ps'}(t)\langle\phi_j^s(t)|V_j|\phi_j^{p}(t)\rangle-A_j^{sp}(t)\langle\phi_j^p(t)|V_j|\phi_j^{s'}(t)\rangle)\nonumber\\
    &\equiv v_j^{ss'}(t)+\frac{\Omega_j}{2\epsilon_j}\sum_p[v_j^{sp}(t)A_j^{ps'}(t)-A_j^{sp}(t)v_j^{ps'}(t)]\,,
\end{align}
and the corresponding Fourier components:
\begin{align}
    V_j^{ss'}(q)\approx v_j^{ss'}(q)+\frac{\Omega_j}{2\epsilon_j}\sum_{p,k}[v_j^{sp}(k)A_j^{ps'}(k-q)-A_j^{sp}(k)v_j^{ps'}(k-q)]\,.
\end{align}
Using these considerations, the leading contributions in $\Omega_j$ gives
\begin{align}
b^{z}_{j}(t)&
\approx-\frac{\Omega_j}{(2\epsilon_j)^2-\omega_0^2}\sum_{s'\in high}(-1)^s\left[\frac{(2\epsilon_j)^2+\omega_0^2}{(2\epsilon_j)^2-\omega_0^2}\left(\frac{i}{\Omega_j}\left(v^{ss'}_j(t)\dot{v}^{s's}_j(t)-\dot{v}^{ss'}_j(t)v^{s's}_j(t)\right)+v^{ss'}_j(t)v^{s's}_j(t)\frac{\gamma_j^s-\gamma_j^{s'}}{\pi}\right)\right.\nonumber\\
&\left.-2v^{0}_j(t)(v_j^{ss'}(t)A_j^{s's}(t)+A_j^{ss'}(t)v_j^{s's}(t))\right]\,,
\end{align}
which is the Floquet analogue of Eq.~\eqref{AdFull}, with  each of the term above having its adiabatic counterpart (in the order presented).

Finally, the exchange coupling becomes:
\begin{align}
J_{1-2}^z(t)&\approx\frac{\omega_0\Omega_1\Omega_2}{2\epsilon_1\epsilon_2}\sum_{s,p\in low ,e,r\in high}(-1)^{s+p}e^{i(q\Omega_j+q'\Omega_{\bar{j}})t}\frac{1}{\omega_0^2-( q\Omega_j)^2}\nonumber\\
&[v_j^{se}(k)A_j^{es}(k-q)+A_j^{se}(k)v_j^{es}(k-q)][v_{\bar{j}}^{pr}(k')A_{\bar{j}}^{rp}(k'-q')+A_{\bar{j}}^{pr}(k')v_{\bar{j}}^{rp}(k'-q')]\,,\\
J_{1-2}^\perp(t)&\approx\frac{\omega_0\Omega_1\Omega_2}{2\epsilon_1\epsilon_2}\sum_{e,r\in high}e^{i(q\Omega_j+q'\Omega_{\bar{j}})t}\frac{2\omega_0}{\omega_0^2-(\gamma_j^1-\gamma_j^{2}-2\pi q)^2\,(\Omega_j/2\pi)^2}\nonumber\\
&[v_j^{1e}(k)A_j^{e2}(k-q)+A_j^{1e}(k)v_j^{e2}(k-q)][v_{\bar{j}}^{2r}(k')A_{\bar{j}}^{r1}(k'-q')+A_{\bar{j}}^{2r}(k')v_{\bar{j}}^{r1}(k'-q')]
\end{align}
which,  in the long time limit and assuming the two frequencies $\Omega_{1,2}$  as being  incommensurate, allows us to keep in the above expression only the $q=q'=0$.

\section{Circular driving}

 Here we provide details for the circular driving case,
 which allows us to map the time-dependent problem to a static one that is amenable to approximations. In Eq.~\eqref{ham_tot} we use the
 electric field parametrization ${\bs n}_j(t)=\{-\sin\theta_j\sin\Omega_j t,\sin\theta_j\cos\Omega_j t,\cos\theta_j \}$, where $\theta_j$ is the cone angle of the $j$-th spin's trajectory, for which the exact solution for bare spin part can be constructed.  Namely, we found that  transformation $U_1(t)\otimes U_2(t)$
 where
 \begin{equation}
 \begin{split}
 U_j(t)=&\frac{1}{\sqrt{2}}\Big(\mathbb{1}_j\!+\!i\sum_{
\alpha}n_{j\alpha}(t)\Gamma^{\alpha 5}_j\Big) {\rm e}^{-i\Omega_j  \Gamma^{12}_j\,t/2} , 
 \end{split}
\end{equation}
with  $\Gamma^{ab}=[\Gamma^a,\Gamma^b]/2i$,
 rotates Hamiltonian to the instantaneous eigenbasis and leaves the remaining  gauge field ($-i U_j(t)^\dagger\dot U_j(t)$) time independent. The resulting Hamiltonian can be further diagonalized with $D_1\otimes D_2$,
\begin{equation}
  \begin{split}
   D_j&=
   \begin{pmatrix}
     \sin\!\frac{\theta_j}{2}\xi_{j+}^+& \cos\!\frac{\theta_j}{2}\xi_{j+}^- & -\cos\!\frac{\theta_j}{2}\xi_{j-}^-& -\sin\!\frac{\theta_j}{2}\xi_{j-}^+ \\
-i\cos\!\frac{\theta_j}{2}\xi_{j+}^+& i\sin\!\frac{\theta_j}{2}\xi_{j+}^- &  -i\sin\!\frac{\theta_j}{2}\xi_{j-}^-&i\cos\!\frac{\theta_j}{2}\xi_{j-}^+ \\
-i\cos\!\frac{\theta_j}{2}\xi_{j-}^+& i\sin\!\frac{\theta_j}{2}\xi_{j-}^- & i\sin\!\frac{\theta_j}{2}\xi_{j+}^-&-i\cos\!\frac{\theta_j}{2}\xi_{j+}^+ \\
  \sin\!\frac{\theta_j}{2}\xi_{j-}^+& \cos\!\frac{\theta_j}{2}\xi_{j-}^-  & \cos\!\frac{\theta_j}{2}\xi_{j+}^-&\sin\!\frac{\theta_j}{2}\xi_{j+}^+\\ 
   \end{pmatrix}\\
   &\xi_{j,s_1=\pm}^{s_2=\pm}=\sqrt{\frac{1}{2}\Big(1+s_1\frac{2\epsilon_j +s_2\,\Omega_j  \cos \theta_j }{2\mathcal{E}_{j,s_2}}\Big)}\\
   &\mathcal{E}_{j,\pm}=\frac{1}{2}\sqrt{\Omega_j^2+4\epsilon_j^2 \pm\,4\Omega_j\epsilon_j\cos\theta_j}.
  \end{split}
 \end{equation} 
Summarizing, the spin-photon Hamiltonian in the rotated frame  $\tilde U(t)=D_1U_1(t)\otimes D_2U_2(t)$  reads,  
\begin{align}
  &\tilde{H}(t) =  \tilde U^\dagger \Big(H_{tot}(t)-i\frac{d}{dt}\Big)\tilde U=
\sum_j\Big(\mathcal{\bar E}_j\Gamma^{5}_j + \delta\mathcal{E}_j\Gamma^{12}_j + g_j\tilde{H}^j_{int}(t)(a^\dagger + a)\Big) + \omega_0\,a^\dagger a\,, \\
  & \tilde{H}^j_{int}(t) = \frac{1}{2}  \sum_{s=\pm}  \Big(  x_{js}(t)(\Gamma^{5}_j +     s\Gamma^{12}_j) + y^R_{js}(t)(\Gamma^{2}_j - s\Gamma^{15}_j)-y^I_{js}(t)(\Gamma^{25}_j + s\Gamma^{1}_j)\Big)\,,
  \label{HSWT}
 \end{align}
where $\mathcal{\bar E}_j=(\mathcal{E}_{j+}+\mathcal{E}_{j-})/2$, $\delta\mathcal{E}_j=(\mathcal{E}_{j+}-\mathcal{E}_{j-})/2$ and $\{g_{jx},g_{jy},g_{jz}\}\equiv g_j \{n^c_{jx},n^c_{jy},n^c_{jz}\}$, $|{\bf n}_j^c|=1$.
Now the entire time-dependence of $\tilde{H}(t)$ is shifted to the spin-photon interaction term through 
 \begin{equation}
     \begin{split}
    &x_{j\pm}(t)=\frac{n^c_{jz}(2\epsilon_j\cos\theta_j\pm\Omega_j)+2\epsilon_j\sin\theta_j(n^c_{jy}\cos\Omega_j t-n^c_{jx} \sin\Omega_j t)}{2\mathcal{E}_{j\pm}}\,,\\ 
     &y^R_{j\pm}(t)=\pm\frac{(\Omega_j\pm 2\epsilon_j\cos\theta_j)(n^c_{jx}\sin\Omega_j t-n^c_{jy}\cos\Omega_j t) \pm 2 n^c_{jz}\epsilon_j\sin\theta_j }{2\mathcal{E}_{j\pm}}\,,\\
     &y^I_{j\pm}(t)=\mp (n^c_{jy}\sin\Omega_j t+n^c_{jx}\cos\Omega_j t),\\ 
     &y_{j\pm}(t)=y^R_{j\pm}(t)+i y^I_{j\pm}(t)\,.
     \end{split}
 \end{equation}

In the  dispersive regime, $g_j\ll \mathcal{E}_{j\pm}$, we can treat the spin-photon coupling in perturbation theory.   we perform the second-order time-dependent Schrieffer-Wolff transformation  (SWT) generated by $\mathcal{A}(t)=\sum_j g_j[a\mathcal{A}_j^+(t)-a^\dagger\mathcal{A}_j^-(t)]$,
\begin{equation}
\mathcal{H}={\rm e}^{\mathcal{A}(t)}\tilde{H}{\rm e}^{-\mathcal{A}(t)}\simeq \omega_0\,a^\dagger a+\sum_j (\mathcal{\bar E}_j\Gamma^{5}_j + \delta\mathcal{E}_j\Gamma^{12}_j)+\frac{1}{2}[\mathcal{A}(t),\sum_jg_j(a^\dagger+a)\tilde{H}^j_{int}(t)]
\end{equation}
 which removes the spin-photon interaction in the leading order if $\mathcal{A}(t)$ satisfies 
 \begin{equation}
  i\mathcal{\dot A}(t)+[\mathcal{A}(t),\omega_0\,a^\dagger a+\sum_j (\mathcal{\bar E}_j\Gamma^{5}_j + \delta\mathcal{E}_j\Gamma^{12}_j)]+{\sum_jg_j(a^\dagger+a)\tilde{H}_{int}^j(t)}=0.  
  \label{SWeq}
 \end{equation}      
 In order to find explicit form of a SWT generator we expand $\mathcal{A}_j^\pm(t)$ and $\tilde{H}_{int}^j(t)$  in the Fourier series (only $n=\{-1,0,1 \}$ coefficients are non-zero),
\begin{equation}
\begin{split}
\mathcal{A}_j^\pm(t)&=\sum_{n=\{-1,0,1\}}\mathcal{A}^\pm_{j,n} {\rm e}^{in\Omega_j t}\\
\tilde{H}^j_{int}&=\sum_{n=\{-1,0,1\}}\tilde{H}_{int}^{j,n}\,{\rm e}^{in\Omega_j t}
\end{split}
\end{equation}
where
\begin{equation}
\begin{split}
\tilde{H}_{int}^{j,n}&=\begin{pmatrix} 
x_{j+}^n&0&0&y_{j+}^{-n*}\\
0&x_{j-}^n &-y_{j-}^{-n*} &0\\
0&-y_{j-}^{n}&-x_{j-}^n&0\\
y_{j+}^{n}&0&0& -x_{j+}^n\\
\end{pmatrix}\\
x_{j\pm}(t)&=\sum_{n=\{-1,0,1\}}x_{j\pm}^n{\rm e}^{in\Omega_j t}\\  
y_{j\pm}(t)&=\sum_{n=\{-1,0,1\}}y_{j\pm}^n{\rm e}^{in\Omega_j t},
\end{split}
\end{equation}

\begin{equation}
\begin{split}
 x_{j\pm}^0&= \frac{n^c_{jz}(\pm\Omega_j+2\epsilon_j\cos\theta_j)}{2\mathcal{E}_{j\pm}}\\
 x_{j\pm}^1&= \frac{(n^c_{jy}+in^c_{jx})\epsilon_j\sin\theta_j}{2\mathcal{E}_{j\pm}}\\
 x_{j\pm}^{-1}&=   \frac{(n^c_{jy}-in^c_{jx})\epsilon_j\sin\theta_j}{2\mathcal{E}_{j\pm}}\\
 y_{j\pm}^0&=\frac{n^c_{jz}\epsilon_j\sin\theta_j}{\mathcal{E}_{j\pm}}\\
 y_{j\pm}^1&=\mp\frac{(n^c_{jy}+in^c_{jx})(2\mathcal{E}_{j\pm}+\Omega_j\pm 2\epsilon_j\cos\theta_j)}{4\mathcal{E}_{j\pm}}\\
 y_{j\pm}^{-1}&=\mp\frac{(n^c_{jy}-in^c_{jx})(-2\mathcal{E}_{j\pm}+\Omega_j\pm 2 \epsilon_j\cos\theta_j)}{4\mathcal{E}_{j\pm}} 
\end{split}    
\end{equation}

The equations for Fourier coefficients $\mathcal{A}^\pm_{i,n}$ resulting from \eqref{SWeq} reads,
\begin{equation}
    \begin{split}
    &(\omega_0-n\Omega_j)A_{j,n}^++[A_{j,n}^+,(\mathcal{\bar E}_j\Gamma^{5}_j + \delta\mathcal{E}_j\Gamma^{12}_j)]+   \tilde{H}_{int}^{j,n}=0\\
    &(\omega_0+n\Omega_j)A_{j,n}^--[A_{j,n}^-,(\mathcal{\bar E}_j\Gamma^{5}_j + \delta\mathcal{E}_j\Gamma^{12}_j)]+  \tilde{H}_{int}^{j,n}=0
    \end{split}
\end{equation}

The solution of \eqref{SWeq} for Fourier coefficients $\mathcal{A}^\pm_{j,n}$ reads,
\begin{equation}
A_{j,n}^+= \begin{pmatrix}
\frac{- x_{j+}^n}{\omega_0-n\Omega_j}&0&0&\frac{-y_{j+}^{-n*}}{\omega_0-n\Omega_j-2\mathcal{E}_{j+}}\\
0&\frac{- x_{j-}^n}{\omega_0-n\Omega_j}&\frac{y_{j-}^{-n*}}{\omega_0-n\Omega_j-2\mathcal{E}_{j-}}&0 \\
0&\frac{y_{j-}^{n}}{\omega_0-n\Omega_j+2\mathcal{E}_{j-}}&\frac{x_{j-}^n}{\omega_0-n\Omega_j}&0 \\
\frac{-y_{j+}^{n}}{\omega_0-n\Omega_j+2\mathcal{E}_{j+}}&0&0&\frac{x_{j+}^n}{\omega_0-n\Omega_j}
\end{pmatrix},
\end{equation}
and $(A^+_{j,n})^\dagger=A^-_{j,-n}$ due to the anti-hermiticity of the SW operator.  

\begin{figure}[t] 
\centering
\includegraphics[width=0.5\linewidth]{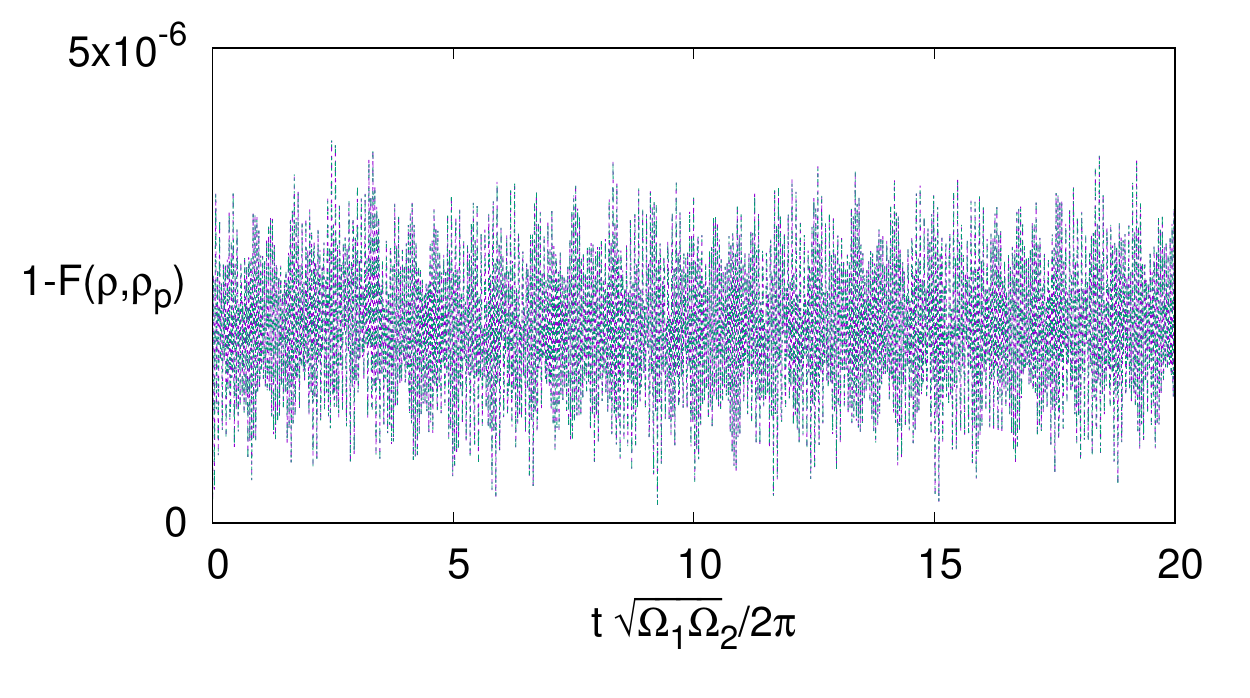} 
\caption{The infidelities $1-F(\tilde \rho_{f1},\rho_{p1})$ and $1-F(\tilde \rho_{f2},\rho_{p2})$ in the adiabatic and dispersive regimes over 20 periods between reduced density matrices obtained by solving  spin part of full  and projected $\mathcal{ H}$ models \eqref{HSWT}. The extremely low infidelity validates the projected Hamiltonian $\mathcal{H}$ for the hole-spin qubit description.}
\label{fidelity} 
\end{figure}

The resulting Hamiltonian within second order in $g_j$ and projected onto the low energy doublet becomes:
\begin{equation}
    \mathcal{H} =\sum_j \mathcal{H}_{j} +\mathcal{H}_{1-2}+\omega_0a^\dagger a+\sum_j\delta\mathcal{H}_{j}\,,\\
\end{equation}
\begin{equation}
\begin{split}
    \mathcal{H}_j&=\Bigg[\frac{\Omega_j \cos\theta_j}{2}-\frac{g_j^2\Omega_j(4\epsilon_j+\omega_0)\sin\theta_j}{4\omega_0(2\epsilon_j+\omega_0)^2}\Big(c_j(t)\sin2\theta_j+b_j(t)\cos2\theta_j \Big)\Bigg] \sigma^z_j\,,\\
    \delta\mathcal{H}_j&=\bigg(\frac{2\epsilon_j g_j^2}{4\epsilon_j^2-\omega_0^2}\big[ 2(n_{jz}^{c\,\, 2}-1)+b_j(t)\cos\theta_j\sin\theta_j+c_j(t)\sin^2\!\theta_j\big] \mathbb{1}^j\\
    &+
    \frac{g_j^2\Omega_j \sin\theta_j(12\epsilon_j^2-\omega_0^2)}{2(4\epsilon_j^2-\omega_0^2)^2}\big[b_j(t)\cos 2\theta_j+c_j(t)\sin 2\theta_j\big] \sigma^z_j\bigg)a^\dagger a\,,\\
    \mathcal{H}_{1-2}&=-\frac{\Omega_1\Omega_2}{2\omega_0}f_1(t)f_2(t)  \sigma^z_1 \sigma^z_2\,,
\end{split}\label{proj}
\end{equation}
where
\begin{equation}
    \begin{split}
        &c_j(t)=1 - 3 n_{jz}^{c\,\, 2} + (n_{jy}^{c\,\,2}-n_{jx}^{c\,\,2}) \cos 2 \Omega_j t - 
  2 n_{jx}^c n_{jy}^c \sin 2 \Omega_j t\,,\\
  &b_j(t)=4n_{jz}^c(n_{jy}^c\cos\Omega_j t-n_{jx}^c\sin\Omega_j t)\,,\\
  &f_j(t)=\Big(n_{jz}^c \sin^2\!\theta_j -\frac{\sin2\theta_j}{4}(n_{jy}^c+in_{jx}^c){\rm e}^{- i \Omega_j t}-\frac{\sin2\theta_j}{4}(n_{jy}^c-in_{jx}^c){\rm e}^{ i \Omega_j t}\Big)g_j/\epsilon_j\,.
    \end{split}
\end{equation}
Here we implicitly considered  $\Omega_j\ll\epsilon_j$ (adiabatic limit), and retained only the terms that are leading order in $\Omega_j$. The above expressions are   showed in the main text.

\subsection{Fidelity and geometrical origin of the entanglement}

In the following we demonstrate that the effective  Hamiltonian restricted to the low-energy sector is indeed representative for a dynamics of the whole system in the adiabatic and dispersive regime. We numerically solve the time-dependent Schr\"odinger equations  for the spin part disregarding the feedback of the photons ($\delta\mathcal{H}_{j}$), for the full $\mathcal{ H}$ (with initial spin wave function $|\psi_j(0)\rangle=\{0,0,\sqrt{1-\beta_j^2},\beta_j{\rm e}^{i\phi_j}\}^T$, $\beta_j\in\{0,1\}$ and $\phi_j\in\{0,2\pi\}$)  and projected   $P_l\mathcal{H}P_l$ ($|\psi_j(0)\rangle=\{\sqrt{1-\beta_j^2}, \beta_j{\rm e}^{i\phi_j}\}^T$) models and obtain two-spin density matrices, $\rho_f(t)$ and $\rho_p(t)$ respectively. 
Next, we project $\tilde\rho_f=P_l\rho_fP_l$ onto low-energy sector and
calculate reduced density matrices for each spin, $\tilde \rho_{fj}$ and $\rho_{pj}$ out of  $\tilde \rho_f$ and $\rho_p$. Finally, we calculate the fidelities, $F(\rho_1,\rho_2)=(\rm{Tr}\sqrt{\sqrt{\rho_1}\rho_2\sqrt{\rho_1}})^2$ between them. In 
Fig. \ref{fidelity} we plot the infidelities  $1-F(\tilde \rho_{f1},\rho_{p1})$ and $1-F(\tilde \rho_{f2},\rho_{p2})$ over 20 mean periods ($20\cdot 2\pi/ \sqrt{\Omega_1\Omega_2}$).  In Fig. \ref{fidelity} we assume the geometry of the cavity set by ${\bs g}_1=g_1\{0.5,0.5,1/\sqrt{2}\}$ and ${\bs g}_2=g_2\{1/\sqrt{2},0.5,0.5\}$ whereas the rest of parameters are chosen as, $\omega_0=0.15$, $\epsilon_1=1.05$, $\epsilon_2=0.95$, $\theta_1=\pi/3$, $\theta_2=\pi/4$,  $g_1=g_2=0.02$, 
$\Omega_1=0.1$, $\Omega_2=0.1/\sqrt{2}$ and hole-spin qubits initial states at $t=0$ are parametrized by $\beta_1=0.3$, $\beta_2=0.4$, $\phi_1=0.7$, $\phi_2=0.4$.
Closeness of infidelities to zero validates the projected model  $\mathcal{H}$.

Next we show that the entanglement, quantified by the concurrence, has geometrical underpinnings even though the interaction Hamiltonian $H_{1-2}$ stems from a product of two velocities, i.e. $H_{1-2}\propto\Omega_1\Omega_2$. In particular, when both spins are driven by incommensurate frequencies, the generation of entanglement is determined by the zeroth Fourier component of Berry curvatures originating from each of the drives, while the other components are averaging out. Such an emergent geometrical dependence means that in the realistic set-up, i.e. in a presence of various sources of noise, the entanglement is more robust than in cases when the dynamical phases are relevant, in particular when the two driving frequencies are incommensurate. Without loss of generality,  let us  assume both spins are driven by classical electric fields following circular trajectories around arbitrary axes (with respect to the direction of the cavity field), and which is described by the Hamiltonian in Eq.~\eqref{proj}.

\begin{figure}[t] 
\centering
\includegraphics[width=0.49\textwidth]{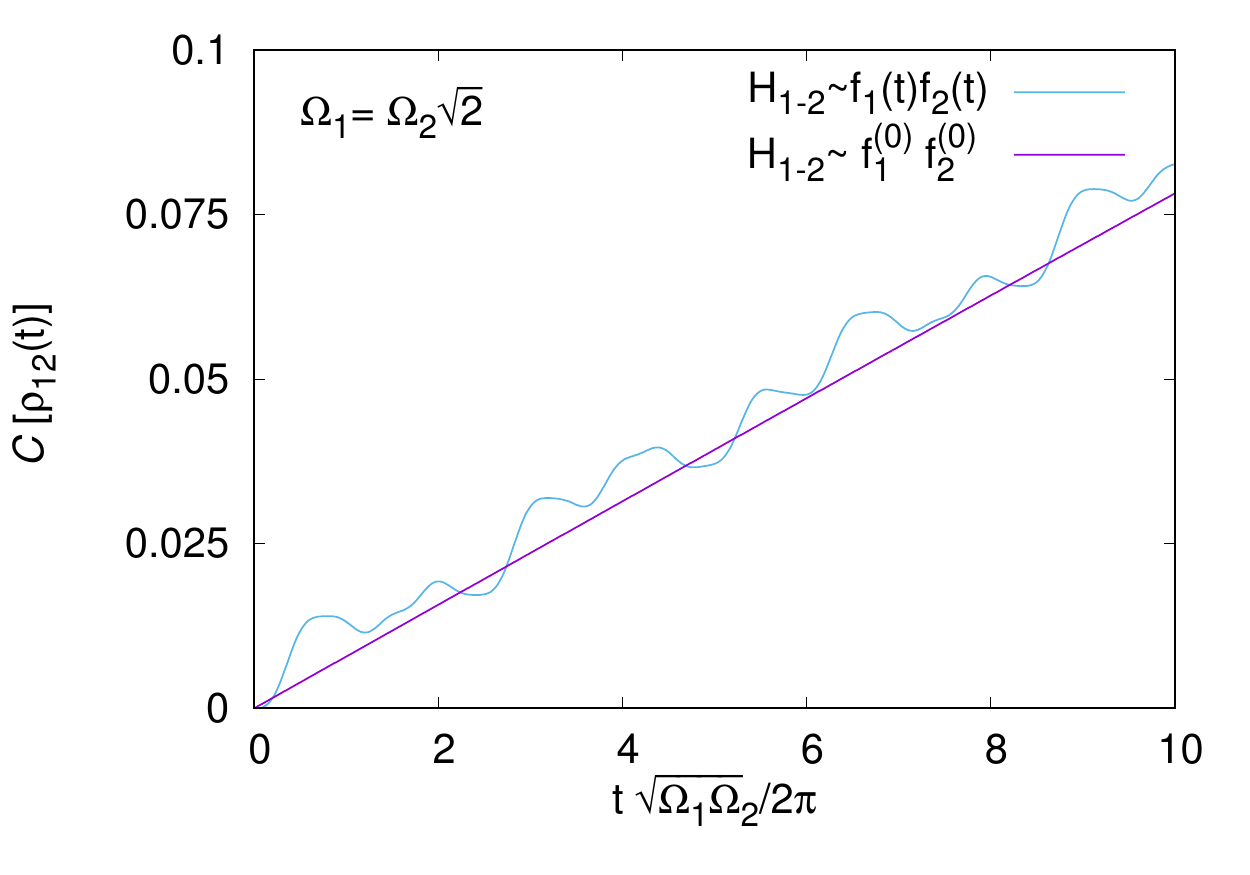} 
\includegraphics[width=0.49\textwidth]{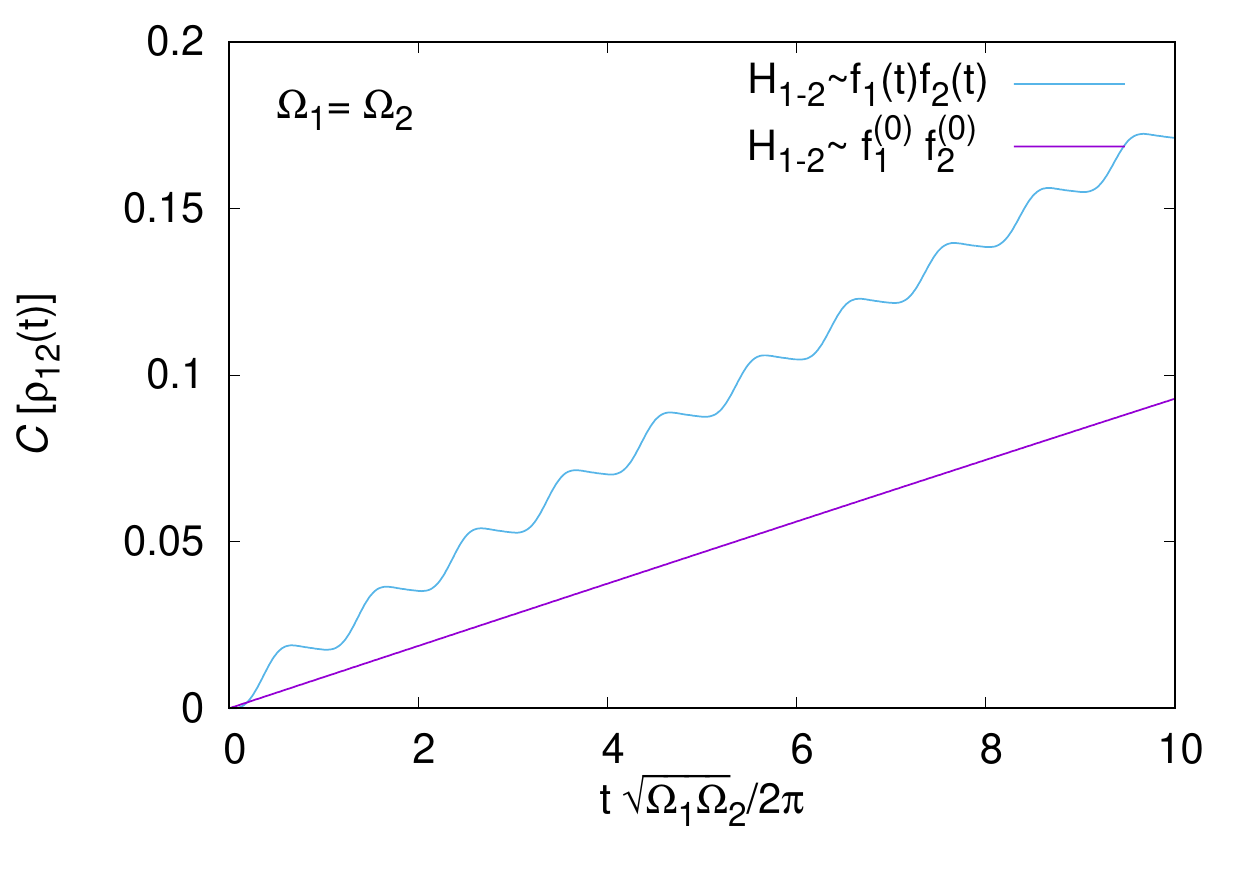}
\caption{The concurrence pertaining to the two-qubit density matrix as a function of time for the full time dependent spin-spin interaction (blue) and for the interaction built from the zeroth Fourier component of the Berry curvature in Floquet basis (purple), i.e. for $f_j(t)\simeq f_j^{(0)}\equiv n_{jz}^c \sin^2\,\theta_j$ in the Hamiltonian \eqref{proj}. In the left panel we show the scenario in which the driving frequencies    are incommensurate,i.e. $\Omega_1=\Omega_2\sqrt{2}$ while in the right panel are commensurate $\Omega_1=\Omega_2$. The spin-photon coupling geometry  is set by ${\bf g}_1=g\{1/2,1/2,1/\sqrt{2}\}$, ${\bf g}_2=g\{1/\sqrt{2},1/2,1/2\}$ and the other parameters taken as $\omega_0=0.15$, $g=0.2$, $\epsilon_1=1.05$, $\epsilon_2=0.95$, $\Omega_1=0.1$, $\beta_1=0.4$, $\beta_2=0.3$, $\theta_1=\pi/3$, $\theta_2=\pi/4$.}
\label{newfig} 
\end{figure}
 
 In Fig. \ref{newfig} we show the entanglement, quantified by concurrence, generated by the full time dependent interaction $H_{1-2}$, Eq.\eqref{proj}  and by the interaction built only from the zeroth Fourier component of the Berry curvature in Floquet basis, i.e. for
 \begin{equation}
 f_j(t)\rightarrow f_j^{(0)}\equiv\frac{1}{T_j}\int_0^{T_j}f_j(t)d\tau =g_jn_{jz}^c \sin^2\,\theta_j/\epsilon_j\,.
 \end{equation}
 Note that since $f_j(t)\propto\dot{\phi}_j,\dot{\theta}_j$, $f_j^{(0)}$ is purely geometrical and thus does not depend on $T_j$ itself.  It is clear that for incommensurate driving frequencies (left panel in Fig. \ref{newfig}) the slope of the concurrence is exactly reproduced by the model that takes into account only zeroth Fourier component of Berry curvature in Floquet basis entering spin-spin interaction, as the effect of the remaining components averages out over time. On the other hand when driving frequencies are equal, $\Omega_1=\Omega_2$, the deviation is substantial, as can be seen in the right panel in Fig.~\ref{newfig}.  For particular case of circular driving, commensurate but unequal  driving frequencies coincidentally provide situation in which exact trend is reproduced by interaction reduced to $H_{1-2}\sim f_1^{(0)}f_2^{(0)}$ because full $H_{1-2}$ has only up to three Fourier components. However, for generic trajectory commensurate driving frequencies provides behavior analogous to equal frequencies in  case of presented here circular driving.     
 
\subsection{Robustness against noise}
 
In the following we analyze robustness of the spin-spin entanglement generation to noise,  in a similar manner as in Ref. \onlinecite{MartinPRX17}. There,  this is  accounted for at the level of classical electric field drives as:
\begin{equation}
    \Omega_j t \rightarrow \phi_j(t)\equiv \Omega_j t + \delta_j(t)\,, 
\end{equation}
where $\delta_j(t)$, with $j=1,2$, is the noise term with a correlation function given by
\begin{equation}
    \langle \dot{\delta}_j(t)\dot{\delta}_j(t') \rangle = \gamma_j(|t-t'|) = \frac{2 \eta_j}{\tau_j\sqrt{2 \pi}}e^{-\frac{|t-t'|^2}{2\tau_j^2}}\,,
\end{equation}
with $\eta_j$ being the noise intensity fulfilling $\eta_j = \int_{0}^{\infty}\gamma_j(t) dt$ and $\tau_j$ being a correlation time giving white-noise limit as $\tau_j\to0$. After a time $T$ one gets phase variance on the order of $\text{var}\left[\phi_j(t)\right] = \eta_j T$ \cite{Romero1999}.

\begin{figure}[t] 
\centering
\includegraphics[width=0.5\linewidth]{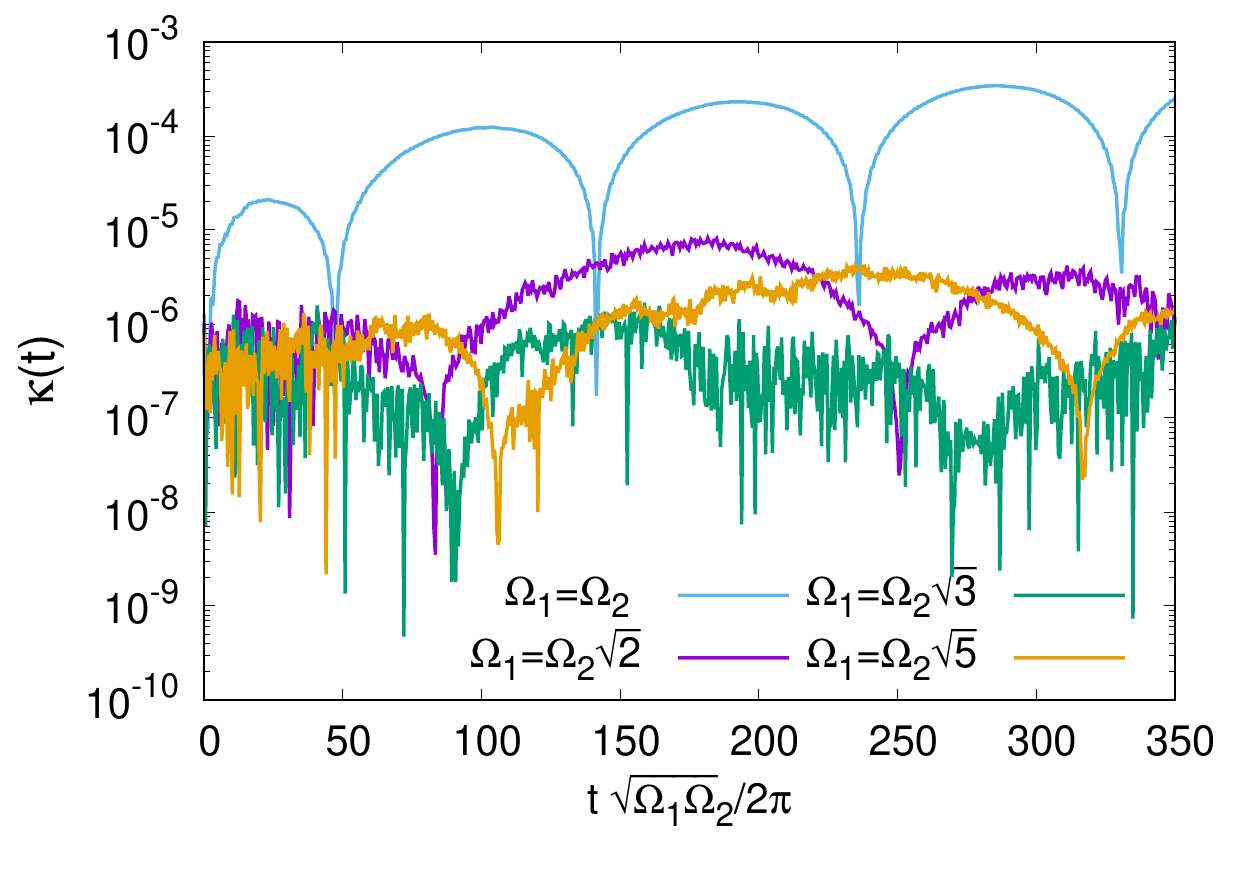} 
\caption{Robustness of the concurrence to decoherence with $\tau_1 = \tau_2 = 100$ and $\eta_1 = \eta_2 = 50$. For the commensurate frequencies $\Omega_2/\Omega_1 = 1$  there is no geometrical protection of the entanglement generation, contrary to the incommensurate frequencies. Cavity coupling geometry  is set by ${\bf g}_1=g\{1/2,1/2,1/\sqrt{2}\}$, ${\bf g}_2=g\{1/\sqrt{2},1/2,1/2\}$ and the other parameters taken as $\omega_0=0.15$, $g=0.2$, $\epsilon_1=1.05$, $\epsilon_2=0.95$, $\Omega_1=0.1$, $\beta_1=0.4$, $\beta_2=0.3$, $\theta_1=\pi/3$, $\theta_2=\pi/4$.}
\label{concurrence_with_noise} 
\end{figure}

We study entanglement generation for the circular driving   in the presence of the noise $\delta_j(t)$ for cavity geometry set by ${\bf g}_1=g\{1/2,1/2,1/\sqrt{2}\}$ and ${\bf g}_2=g\{1/\sqrt{2},1/2,1/2\}$ .
In Fig.~\ref{concurrence_with_noise} we plot $\kappa = |{\cal C}_0(t) - \overline{{\cal C}(t)}|$, where ${\cal C}_0(t)$ is noiseless concurrence and $\overline{{\cal C}(t)}$ is a mean concurrence averaged over 20 different realization of the Wienner process $\delta_j(t)$. 
For $\varOmega_1 = \varOmega_2$, $\kappa(t)$ clearly increases over time, contrary to the case of $\varOmega_1 \ne \varOmega_2$ indicating robustness of the provided protocol to external noise.  

\endwidetext

\end{document}